\documentclass[prd,amsfonts,aps,nofootinbib,notitlepage,11pt,superscriptaddress,floatfix]{revtex4-1}

\pdfoutput=1
\usepackage{amsmath,amssymb}
\usepackage{graphicx}
\usepackage[utf8]{inputenc}
\usepackage{cancel}
\usepackage[colorlinks=true]{hyperref}
\usepackage{color}
\usepackage{array,multirow}

\usepackage{float}

\newcommand{\UPTC}{Escuela de Física, Universidad Pedagógica y Tecnológica de Colombia,\\
Avenida Central del Norte \# 39-115, Tunja, Colombia}
\newcommand{\UdeA}{Instituto de Física, Universidad de Antioquia,\\Calle 70 \# 52-21, Apartado Aéreo 1226, Medellín, Colombia}

\begin{document}
\title{Two-component scalar dark matter in $Z_{2n}$ scenarios}

\author{Carlos E. Yaguna}
\affiliation{\UPTC}
\author{\'Oscar Zapata}
\affiliation{\UdeA}

\begin{abstract}
In multi-component scalar dark matter scenarios, a single $Z_N$ ($N\geq 4$) symmetry may account for the stability of different dark matter particles.  Here we study the case where $N$ is even ($N=2n$) and two species, a complex scalar and a real scalar,  contribute to the observed dark matter density.  We perform a phenomenological analysis of three scenarios based on the $Z_4$ and $Z_6$ symmetries, characterizing their viable parameter spaces and analyzing their detection prospects. Our results show that, thanks to the new interactions allowed by the $Z_{2n}$ symmetry, current experimental constraints can be satisfied over a wider range of dark matter masses, and that these scenarios may lead to observable signals in direct detection experiments. Finally,  we argue that these three scenarios serve as prototypes for other two-component  $Z_{2n}$ models with one complex and one real dark matter particle. 
  \end{abstract}

\maketitle
\section{Introduction}
Among the long series of  models that have been proposed to explain the dark matter, the scalar singlet model~\cite{Silveira:1985rk,McDonald:1993ex,Burgess:2000yq} constitutes the simplest realization where a single particle lying around the electroweak scale has the suitable self-annihilation rate to achieve the observed dark matter abundance. Nevertheless, it  is severely constrained by current observations.  The dark matter mass must lie, in fact, either at  the Higgs-resonance or above the TeV scale~\cite{Cline:2013gha,Athron:2018ipf}.  Appealing alternatives to this paradigm are multi-component dark matter scenarios featuring scalar singlet fields that are stabilized by a  single $Z_N$  ($N\geq 4$) symmetry~\cite{Batell:2010bp,Belanger:2012vp,Belanger:2014bga,Yaguna:2019cvp}\footnote{See Refs.~~\cite{Boehm:2003ha,Ma:2006uv,Cao:2007fy,Hur:2007ur,Lee:2008pc,Zurek:2008qg,Barger:2008jx,Profumo:2009tb,Belanger:2011ww,Baer:2011hx,Liu:2011aa,Ivanov:2012hc,Belanger:2014vza,Esch:2014jpa} for pioneering studies on multi-component dark matter scenarios.}.

The recent phenomenological study of the $Z_5$-invariant two-component dark matter model ~\cite{Belanger:2020hyh}, for instance, revealed that the dark matter masses may actually lie below the TeV while being consistent with current experimental bounds, and that  both dark matter particles may give rise to observable signals in  ongoing and forthcoming direct dark matter detection experiments. This $Z_5$ model, in addition,  serves as a prototype for all the  two-component scenarios based on a $Z_N$ symmetry where the dark matter particles are both complex scalars. In this paper we will focus instead on models where the dark matter consists of a complex scalar and a real scalar. 

This type of two-component scalar dark matter arises when $N$ is even and one of the fields transforms as $\phi\to -\phi$ under the $Z_N$. In this work we explicitly consider the models based on the $Z_4$ and $Z_6$ symmetries.  The latter admits two different charge assignments for the dark matter fields, one of them featuring  unconditional stability \cite{Yaguna:2019cvp} --an unusual property among $Z_N$ models.  For these three scenarios we perform a phenomenological analysis with the aim of identifying and characterizing the viable parameter space.   Notably, we find that the dark matter masses may lie below the TeV and that the predicted direct detection rates can be within the expected sensitivity of future  experiments~\cite{Akerib:2018lyp,Aalbers:2016jon}. This new class of two-component dark matter models is thus shown to be not only theoretically compelling and phenomenologically consistent, but also experimentally testable.

The rest of the paper is organized as follows. In Section \ref{sec:CPV} we present a general discussion of two-component dark matter scenarios under a $Z_{2n}$ symmetry, and introduce  the $Z_4$ and $Z_6$ models that are the focus of our analysis. The phenomenological analysis is presented in \ref{sec:pheno}, where the viable parameter space of each scenario is established and illustrated, along with its detection prospects. A possible extension of these scenarios is introduced in Section \ref{sec:extension} while a generalization of our results to other $Z_{2n}$ symmetries is sketched in Section  \ref{sec:discussion}.  Finally, in Section \ref{sec:conclusions} we draw our conclusions. 

\section{$Z_{2n}$ models}\label{sec:CPV}
The two-component dark matter scenarios based on a $Z_N$ symmetry  extend the Standard Model (SM) with  two  scalar fields, $\phi_A$ and $\phi_B$, that are SM singlets and  have  different $Z_N$ charges,  $w_A$ and $w_B$ respectively\footnote{Two-component scenarios with scalar singlets (doublets) stabilized by two $Z_2$ or $Z_3$ symmetries have been studied in Refs.~\cite{Modak:2013jya,Biswas:2015sva,Bhattacharya:2016ysw,Bhattacharya:2017fid,Pandey:2017quk} (\cite{Hernandez-Sanchez:2020aop}).}. If $N$ is odd,  these fields are necessarily complex and they both contribute to the observed dark matter density.  The phenomenology of these scenarios was studied in detail in Ref.~\cite{Belanger:2020hyh}. Here, we will focus instead in the case where $N$ is even  and one of the fields -say $\phi_B$- is real with a charge $w_B=w^{N/2}=-1$, since $w=e^{i2\pi/N}$. 
Consequently, in these scenarios the dark matter consists of a complex scalar field ($\phi_A$) and a real scalar field ($\phi_B$).

Let us consider the most general scalar potential invariant under a $Z_{2n}$ symmetry for two scalar singlets, one complex and one real, with $\phi_B$ having a $Z_{2n}$ charge $w^n=-1$. The corresponding potential can be written as the sum of two contributions,
\begin{align}\label{eq:Z2nlag}
 \mathcal{V}_{Z_{2n}}(\phi_A,\phi_B)&=\,\mathcal{V}_{1}(\phi_A,\phi_B)+\mathcal{V}_2(\phi_A,\phi_B). 
 \end{align}
 The first one  corresponds to the terms that are invariant for any $Z_{2n}$ symmetry,
 \begin{align}\label{eq:V1}
 \mathcal{V}_1(\phi_A,\phi_B)&\equiv\,\,\mu^2_H|H|^2+\lambda_H|H|^4+\mu_{A}^2|\phi_A|^2+\lambda_{4A}|\phi_A|^4+\frac{1}{2}\mu_{B}^2\phi_B^2+\lambda_{4B}\phi_B^4\nonumber\\
  & \,+\lambda_{4AB}|\phi_A|^2\phi_B^2+\lambda_{SA}|H|^2|\phi_A|^2+\frac{1}{2}\lambda_{SB}|H|^2\phi_B^2,
 \end{align}
 where $H$ is the SM Higgs doublet. All the parameters appearing in $ \mathcal{V}_1$ are real.

The interactions  present in  $\mathcal{V}_1$ may affect the dark matter phenomenology of the these scenarios in different ways. The quartic interaction $\lambda_{4AB}$, for instance, induces  dark matter conversion processes, while the $\lambda_{Si}$ terms lead to $\phi_i \phi_i$ annihilations and to trilinear interactions between the dark matter particles and  the Higgs. These interactions are crucial because they determine the elastic scattering of the dark matter particles off nuclei, providing a way to test these models via direct detection experiments. Notice that these models  are particular realizations of the Higgs portal scenarios~\cite{Patt:2006fw,Arcadi:2019lka}, and natural extensions of the singlet scalar model~\cite{Silveira:1985rk,McDonald:1993ex,Burgess:2000yq}.

The expression for the spin-independent (SI) cross-section is exactly the same as in the singlet scalar (or Higgs-portal) model and reads
 \begin{align}
 \label{eq:sip}
     \sigma_{\phi_i}^{{\rm SI}}&=\frac{\lambda_{Si}^2}{4\pi}\frac{\mu_R^2 m_p^2 f_p^2}{m_h^4 M_{\phi_i}^2},
 \end{align}
 where $\mu_R$ is the reduced mass, $m_p$ the proton mass, $f_p\approx 0.3$ is the quark content of the proton, and 
 $M_{\phi_i}^2=\mu_{i}^2+\frac{1}{2}\lambda_{Si}v_H^2$ is the mass of the dark matter particle.

The second contribution to the scalar potential, $\mathcal{V}_2$, contains the terms that are invariant for  the specific $Z_{2n}$ symmetry under consideration and for the given set of scalar fields, depending on their charges. Next we will examine $\mathcal{V}_2$ for the simplest realizations of these scenarios --those based on the $Z_4$ and $Z_6$ symmetries.

\subsection{$Z_4$ model}\label{sec:models}

\begin{figure}[t]
\centering
\includegraphics[scale=0.9]{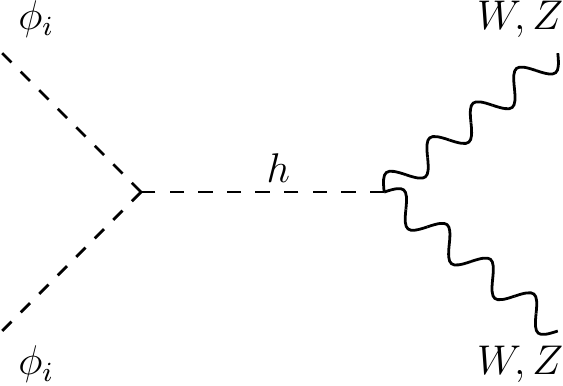}\hspace{0.4cm}
\includegraphics[scale=0.9]{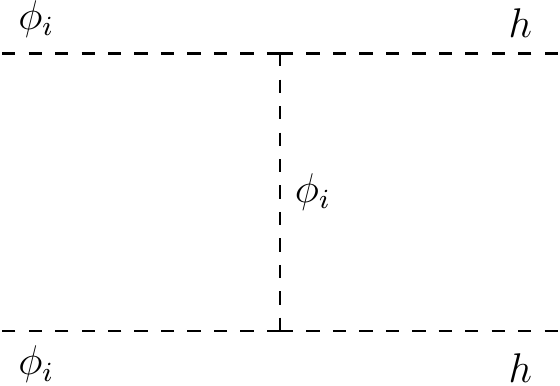}\hspace{0.4cm}
\caption{Dark matter $\phi_i$ annihilation processes involving the $\lambda_{Si}$ interactions. These processes are common to both $Z_4$ and $Z_6$ models.}
\label{fig:coanni}
\end{figure}

\begin{figure}[t]
\centering
\includegraphics[scale=0.9]{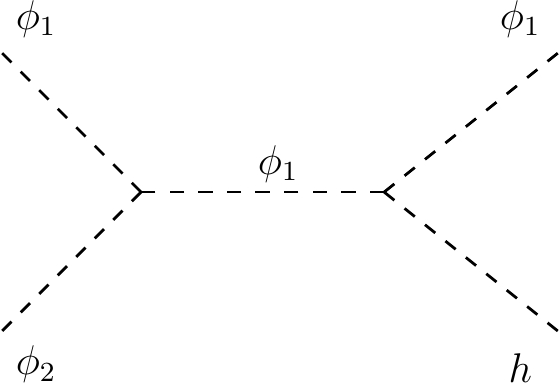}\hspace{0.4cm}
\includegraphics[scale=0.9]{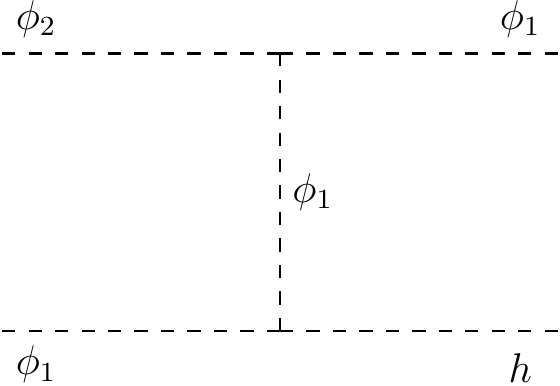}\hspace{0.4cm}
\includegraphics[scale=0.9]{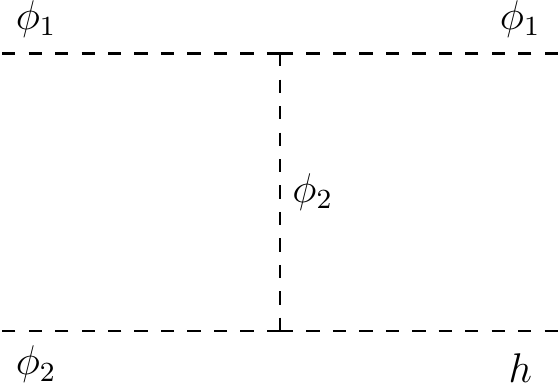}\\
\vspace{0.4cm}
\includegraphics[scale=0.9]{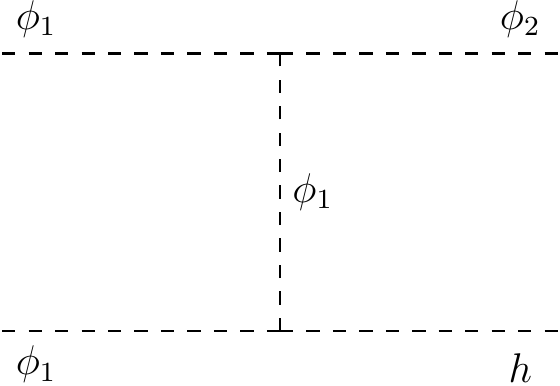}\hspace{1cm}
\includegraphics[scale=0.9]{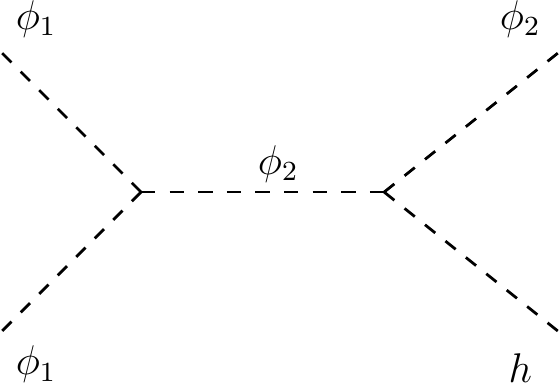}
\caption{Semi-annihilation processes in the $Z_4$ model, which are generated by the $\mu_{S1}$ interaction along with one Higgs portal interaction. }
\label{fig:semi2}
\end{figure}

In the $Z_4$ model the two scalar fields transform as 
 \begin{align}
&    \phi_1\sim \omega_4,\,\,\, \phi_2\sim \omega_4^2,             
\end{align}
with $\omega_4=\exp(i2\pi/4)$ and  $\phi_2$ playing the role of $\phi_B$ in the previous section. 
The $\mathcal{V}_2$ part of the scalar potential is
\begin{align}\label{eq:V3Z4}
  \mathcal{V}_2^{Z_4}(\phi_1,\phi_2)&=\, \frac{1}{2}\left[\mu_{S1}\phi^2_1\phi_2 + \lambda_{51} \phi_1^4\right]  + \text{h.c.},
\end{align}
which includes trilinear and quartic interactions among the new fields. The latter is a self-interaction term that plays no role in the present phenomenological study. The former, on the contrary, would render $\phi_2$ unstable via the two-body decay $\phi_{2}\to\phi_1+\phi_1$. To obtain a two-component dark matter scenario it becomes necessary, therefore, to require  $M_{\phi_2}<2M_{\phi_1}$ so that $\phi_2$ remains stable and contributes to the dark matter density. From now on this condition is assumed to be satisfied. The trilinear term, then, leads to dark matter conversion processes, and the interplay with the $\lambda_{S1}$ and $\lambda_{S2}$ interactions gives rise to the semi-annihilation processes showed in Fig. \ref{fig:semi2}. As we will see in the next section,  semi-annihilations play a major role in this model, allowing for a wider range of viable dark matter masses.  Notice that the processes in the top panels modify the  $\phi_2$ number density by  one unit  whereas those in the bottom panels  change the $\phi_1$ number density by two units  and the $\phi_2$ number density by one unit.     

\begin{table}[t]
    
    \begin{tabular}{c |c}
      $\phi_1$ Processes   & Type  \\
      \hline 
        $\phi_1+\phi_1^\dagger\to SM + SM$ & $1100$\\
        $\phi_1+\phi_1^\dagger \to \phi_2+\phi_2$  & $1122$\\
        $\phi_1 + \phi_1 \to \phi_2+h  $ & $1120$\\
    \end{tabular}\hspace{2cm}
    \begin{tabular}{c |c}
      $\phi_2$ Processes   & Type  \\
      \hline 
        $\phi_2+\phi_2\to SM + SM$ & $2200$\\
        $\phi_2+\phi_2 \to \phi_1+\phi_1^\dagger$  & $2211$\\
        $\phi_2+\phi_1 \to \phi_1^\dagger+h$  & $2110$\\
        $ \phi_2 + h\to \phi_1+\phi_1 $ & $2011$\\
    \end{tabular}
    
    \caption{The $2\to 2$ processes that are allowed in the $Z_4$ model and that can modify the relic density of $\phi_1$ (left) and $\phi_2$ (right). $h$ denotes the SM Higgs boson. Conjugate and inverse processes are not shown.  
    }
    \label{tab:processesZ4}
\end{table}

The different processes that contribute to the relic densities of $\phi_1$ and $\phi_2$ are shown in table \ref{tab:processesZ4}, classified according to their type. The Boltzmann equations for the $Z_4$ model thus read
\begin{align}
\label{boltzmann1}
\frac{dn_1}{dt}&=-\sigma_v^{1100}  \left(n_1^2-\bar{n}_1^2 \right) -
\sigma_v^{1120}\left( n_1^2- n_2 \frac{\bar{n}_1^2}{\bar{n}_2} \right)
- \sigma_v^{1122}\left( n_1^2- n_2^2 \frac{\bar{n}_1^2}{\bar{n}_2^2}
\right)  - 3H n_1, \\
\frac{dn_2}{dt}&=-\sigma_v^{2200}  \left(n_2^2-\bar{n}_2^2 \right) 
- \sigma_v^{2211}\left( n_2^2- n_1^2 \frac{\bar{n}_2^2}{\bar{n}_1^2}
\right)-\frac{1}{2}\sigma_v^{1210}\left( n_1 n_2- n_1 \bar{n}_2 \right)+\frac{1}{2}\sigma_v^{1120}(n_1^2-n_2\frac{\bar{n}_1^2}{\bar{n}_2})   - 3H n_2.      
\label{boltzmann2}
\end{align}
Here $\sigma_v^{abcd}$  stands for  the thermally averaged cross section, which satisfies 
\begin{equation}
    \bar{n}_a\bar{n}_b\sigma_v^{abcd}=\bar{n}_c\bar{n}_d\sigma_v^{cdab}.
\end{equation}
whereas $n_{i}$ ($i=1,2$) denote the number densities of $\phi_i$, and $\bar{n}_i$  their respective equilibrium values.  To numerically solve these equations and obtain the relic densities we use {\tt micrOMEGAs}~\cite{Belanger:2014vza} throughout this paper. Since
 its version 4.1, {\tt micrOMEGAs} incorporates two-component dark matter scenarios, automatically including all the relevant processes in a given model.

Even though the free parameters  in the $\mathcal{V}_2$ potentials are in principle complex, taking them to be real is not expected to modify in any significant way the results. For simplicity, in the following it is assumed that they are real. 

\subsection{$Z_6$ model}

\begin{table}[t]
    
    \begin{tabular}{c |c}
      $\phi_1$ Processes   & Type  \\
      \hline 
        $\phi_1+\phi_1^\dagger\to SM + SM$ & $1100$\\
        $\phi_1+\phi_1^\dagger \to \phi_3+\phi_3$  & $1133$\\
        $\phi_1 + \phi_1 \to \phi_3+\phi_1^\dagger  $ & $1131$\\
    \end{tabular}\hspace{2cm}
    \begin{tabular}{c |c}
      $\phi_3$ Processes   & Type  \\
      \hline 
        $\phi_3+\phi_3\to SM + SM$ & $3300$\\
        $\phi_3+\phi_3 \to \phi_1+\phi_1^\dagger$  & $3311$\\
        $\phi_3+\phi_1 \to \phi_1^\dagger+\phi_1^\dagger$  & $3111$\\
    \end{tabular}
    
    \caption{The $2\to 2$ processes that are allowed in the $Z_6(13)$ model and that can modify the relic density of $\phi_1$ (left) and $\phi_3$ (right). Notice that semi-annihilation processes are not allowed in this model.  Conjugate and inverse processes are not shown.  
    }
    \label{tab:processesZ613}
\end{table}

For the  $Z_6$ model there are two possible charge assignments
 \begin{align}
   &    \phi_1\sim \omega_6,\,\,\, \phi_3\sim \omega_6^3, 
\hspace{1cm}\text{or}\hspace{1cm}\phi_2\sim \omega_6^2,\,\,\, \phi_3\sim \omega_6^3,
\end{align}
with $\omega_6=\exp(i\pi/3)$ and $\phi_3$ playing the role of $\phi_B$ in the previous section. Thus,  two distinct scenarios arise, which we denote as $Z_6(13)$ and $Z_6(23)$ respectively.
The scalar potential $\mathcal{V}_2$ for the $Z_6(13)$ scenario turns out to be
\begin{align}\label{eq:V2Z613}
 \mathcal{V}_{2}^{Z_6}(\phi_1,\phi_3)&=\frac{1}{3} \lambda_{41}'\phi _1^3\phi_3 + \text{h.c.},
\end{align} 
and includes only quartic interactions. To guarantee that  $\phi_3$ is stable, the condition $M_{\phi_3}<3M_{\phi_1}$ is assumed to hold in the following.  Due to the absence of trilinear interactions, there are no semi-annihilation processes in this case. The $\lambda_{41}'$ interaction gives rise  to dark matter conversion processes only.  The processes that affect the relic density are displayed in table \ref{tab:processesZ613}. From them, the Boltzmann equations can be written down.

\begin{figure}[t]
\centering
\includegraphics[scale=0.9]{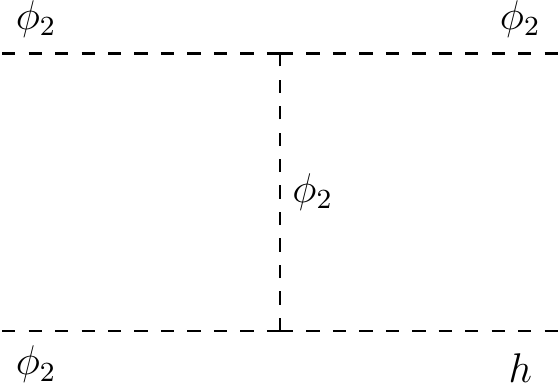}\hspace{1cm}
\includegraphics[scale=0.9]{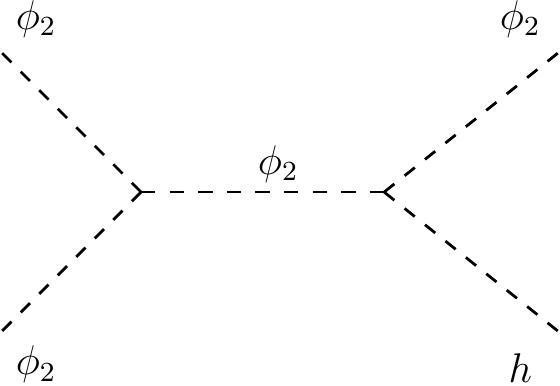}
\caption{Semi-annihilation processes in the $Z_6(23)$ scenario governed by the trilinear self-interaction $\mu_{32}$ and the Higgs portal $\lambda_{S2}$.}
\label{fig:semi3}
\end{figure}

For the $Z_6(23)$ scenario,  $\mathcal{V}_2$ reads 
\begin{align}\label{eq:V2Z623}
 \mathcal{V}_2^{Z_6}(\phi_2,\phi_3)&=\frac{1}{3}\mu_{32}\phi _2^3 +\text{h.c.}.
\end{align}
Since there are neither cubic nor quartic terms involving one single field,  $\phi_2$ and $\phi_{3}$ are both stable independently of their masses --an unusual situation dubbed \emph{unconditional stability} \cite{Yaguna:2019cvp}.  The $Z_6$ symmetry with $\phi_{2}$ and $\phi_{3}$ is, in fact, the simplest realization of unconditional stability for two dark matter particles\footnote{Notice that  unconditional stability is not limited to the renormalizable Lagrangian but is maintained for operators of arbitrary dimension.}. The cubic self-interaction term $\mu_{32}$ in Eq.~(\ref{eq:V2Z623})  along with the $\lambda_{S2}$ interaction give rise to semi-annihilation processes affecting only the number of $\phi_2$ particles (see Figure \ref{fig:semi3}), which, in the next section, will be shown to be quite important.  Table \ref{tab:processesZ623} shows the processes that must be taken into account in the Boltzmann equations for this model.

\begin{table}[t]
    
    \begin{tabular}{c |c}
      $\phi_2$ Processes   & Type  \\
      \hline 
        $\phi_2+\phi_2^\dagger\to SM + SM$ & $2200$\\
        $\phi_2+\phi_2^\dagger \to \phi_3+\phi_3$  & $2233$\\
        $\phi_2 + \phi_2 \to \phi_2^\dagger+h  $ & $2220$\\
    \end{tabular}\hspace{2cm}
    \begin{tabular}{c |c}
      $\phi_3$ Processes   & Type  \\
      \hline 
        $\phi_3+\phi_3\to SM + SM$ & $3300$\\
        $\phi_3+\phi_3 \to \phi_2+\phi_2^\dagger$  & $3322$\\
    \end{tabular}
    
    \caption{The $2\to 2$ processes that are allowed in the $Z_6(23)$ model and that can modify the relic density of $\phi_1$ (left) and $\phi_3$ (right).  $h$ denotes the SM Higgs boson. Conjugate and inverse processes are not shown.  
    }
    \label{tab:processesZ623}
\end{table}

This  $Z_6(23)$ scenario can be seen as a two-component extension of the  $Z_3$ scalar singlet model \cite{Ma:2007gq,Belanger:2012zr,Hektor:2019ote}, for it includes all the terms present in it.  Recently, a $U(1)$ extension of this scenario was studied in Ref.~\cite{Choi:2021yps}, which reaffirms one of the main theoretical advantages of $Z_N$ models: they can be easily incorporated into gauge extensions of the Standard Model. 

\section{The viable parameter space}\label{sec:pheno}
Here  we will obtain and study  viable regions of the $Z_4$ and $Z_6$ models described in the previous section. To that end, the parameter space will be randomly scanned and the points compatible with the constraints obtained from the invisible decays of the Higgs boson , dark matter density~\cite{Aghanim:2018eyx} and direct dark matter searches~\cite{Aprile:2018dbl} will be selected.  This sample of viable points will then be analyzed, paying particular attention to the  dark matter detection prospects.  Let us stress that this random sampling of the parameter space does not allow a statistical interpretation of the distribution of viable points --it cannot be used to identify the best fit point or the most favored regions. It should be enough, however, for our purposes: to find the most relevant parameters and to identify the mechanisms that allow to satisfy the current bounds.  
 
If the dark matter particles are lighter than half the Higgs mass, the decay $h\to \phi_i^*\phi_i$ would be allowed,  contributing to the invisible branching ratio of the Higgs boson ($\mathcal{B}_{inv}$). The decay width associated with $h\to \phi_i^*\phi_i$ is
\begin{align}
    \Gamma(h\to \phi_i^*\phi_i)&=\eta_{\phi_i}\frac{\lambda^2_{Si}v_H^2}{32\pi M_h}\left[1-\frac{4M^2_{\phi_i}}{M_h^2}\right]^{1/2},
\end{align}
where $\eta_{\phi_i}=1\,(2)$ for $\phi_i$ real (complex). To be consistent with current data, we require that $\mathcal{B}_{inv}\leq0.13$) \cite{Sirunyan:2018owy,ATLAS:2020cjb}.

The relic density constraint reads 
\begin{equation}
    \Omega_{\phi_A}+\Omega_{\phi_B}=\Omega_{\text{DM}},
\end{equation}
where $\Omega_{\text{DM}}$ is the dark matter abundance as reported by PLANCK~\cite{Aghanim:2018eyx}, 
\begin{align}
    \Omega_{\text{DM}}h^2=0.1198\pm 0.0012. 
\end{align}
We consider a model to be consistent with this measurement if its relic density, as computed by micrOMEGAs, lies between $0.11$ and $0.13$, which takes into account an estimated theoretical uncertainty of order $10\%$. Since we have two dark matter particles, an important  quantity  in our analysis is the fractional contribution of each to the total dark matter density, $\xi_{\phi_i}\equiv \Omega_{\phi_i}/\Omega_{\text{DM}}$, with $\xi_{\phi_A} +\xi_{\phi_B}=1$.

In our study, we require the spin-independent cross section, computed from equation (\ref{eq:sip}), to be below the direct detection limit set by the XENON1T collaboration \cite{Aprile:2018dbl}. Such direct detection limit usually provides very strong constraints on Higgs-portal scenarios like the models we are discussing.  In particular, for the singlet real scalar  model~\cite{Silveira:1985rk,McDonald:1993ex,Burgess:2000yq} the minimum dark matter mass compatible with upper limit set by the XENON1T collaboration is $\sim950$ GeV (for the complex case turns to be $\sim2$ TeV). As we will show, however, the new interactions present in the $Z_4$ and $Z_6$ scenarios  allow us to simultaneously satisfy the relic density constraint and direct detection limits for lower dark matter masses.

In addition, we will study the testability of the viable  models at future direct detection experiments including  LZ~\cite{Akerib:2018lyp} and DARWIN \cite{Aalbers:2016jon}, as well as the possible constraints from indirect detection searches. For these searches, the relevant particle physics quantity is no longer $\langle\sigma v\rangle$ but $\xi_i\xi_j\langle\sigma v\rangle_{ij}$, where  $\langle\sigma v\rangle_{ij}$ is the  cross section times velocity for the annihilation process of dark matter particles $i$ and $j$ into a certain final state. The computation of the different annihilation rates is performed with the help of  micrOMEGAS. We will rely on the limits and on the projected sensitivities reported by the Fermi collaboration from  $\gamma$-ray observations of dShps  \cite{Ackermann:2015zua,Charles:2016pgz}.

We have performed several random scans, varying just a subset of the free parameters of the model at a time so as to make the analysis simpler. In all the scans, the dark matter masses and the Higgs-portal couplings are randomly sampled (using a logarithmically-uniform distribution) in the following ranges:
\begin{align}
   &40\,{\rm GeV}\leq M_{\phi_A}, M_{\phi_B} \leq 2\,{\rm TeV},\\
   &10^{-4}\leq |\lambda_{SA}|,\,|\lambda_{SB}|\leq 1.
\end{align}
The remaining parameters are crucial because they alleviate the strong correlation  between the dark matter annihilations that set the relic density and  the scattering off nuclei that are constrained by direct detection searches, probably allowing for viable models at low dark matter masses ($M_{\phi_i}\lesssim 1$~TeV). The $Z_5$ model \cite{Belanger:2020hyh}, for instance, became viable over the entire dark matter mass range due to the trilinear interaction term $\phi_1^2\phi_2$. We expect, therefore, a similar situation in the $Z_4$ model and, up to a large extent, in the $Z_6(23)$ scenario. The dimensionful parameters, $\mu_{S1}$ and $\mu_{32}$, will be varied (within each scenario) up to a maximum value (10 TeV), which is not far from the maximum value allowed for the heaviest dark matter particle ($4$ TeV).

\subsection{$Z_4$ model }
\begin{figure}[t]
\centering
\includegraphics[scale=0.44]{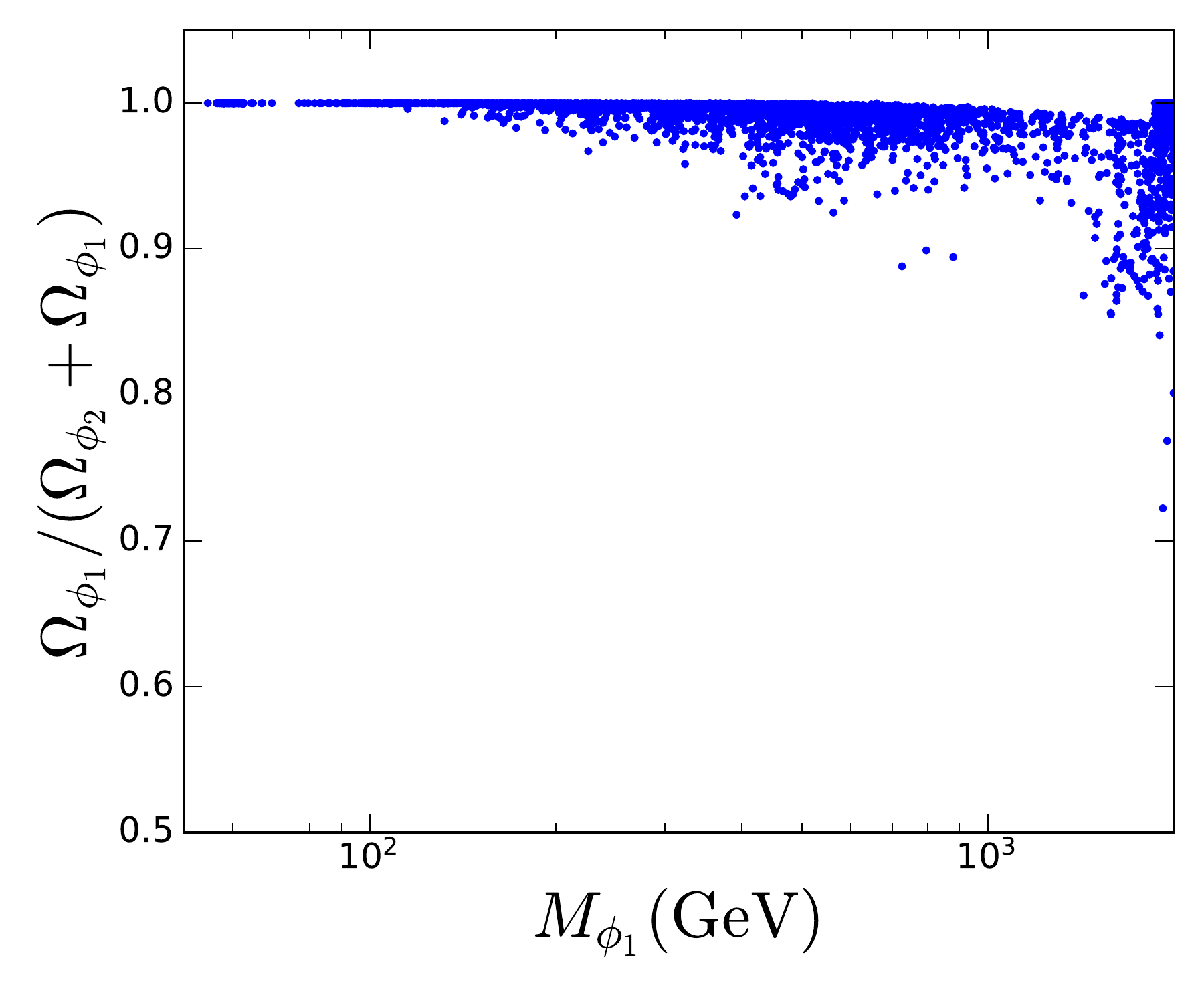}
\includegraphics[scale=0.44]{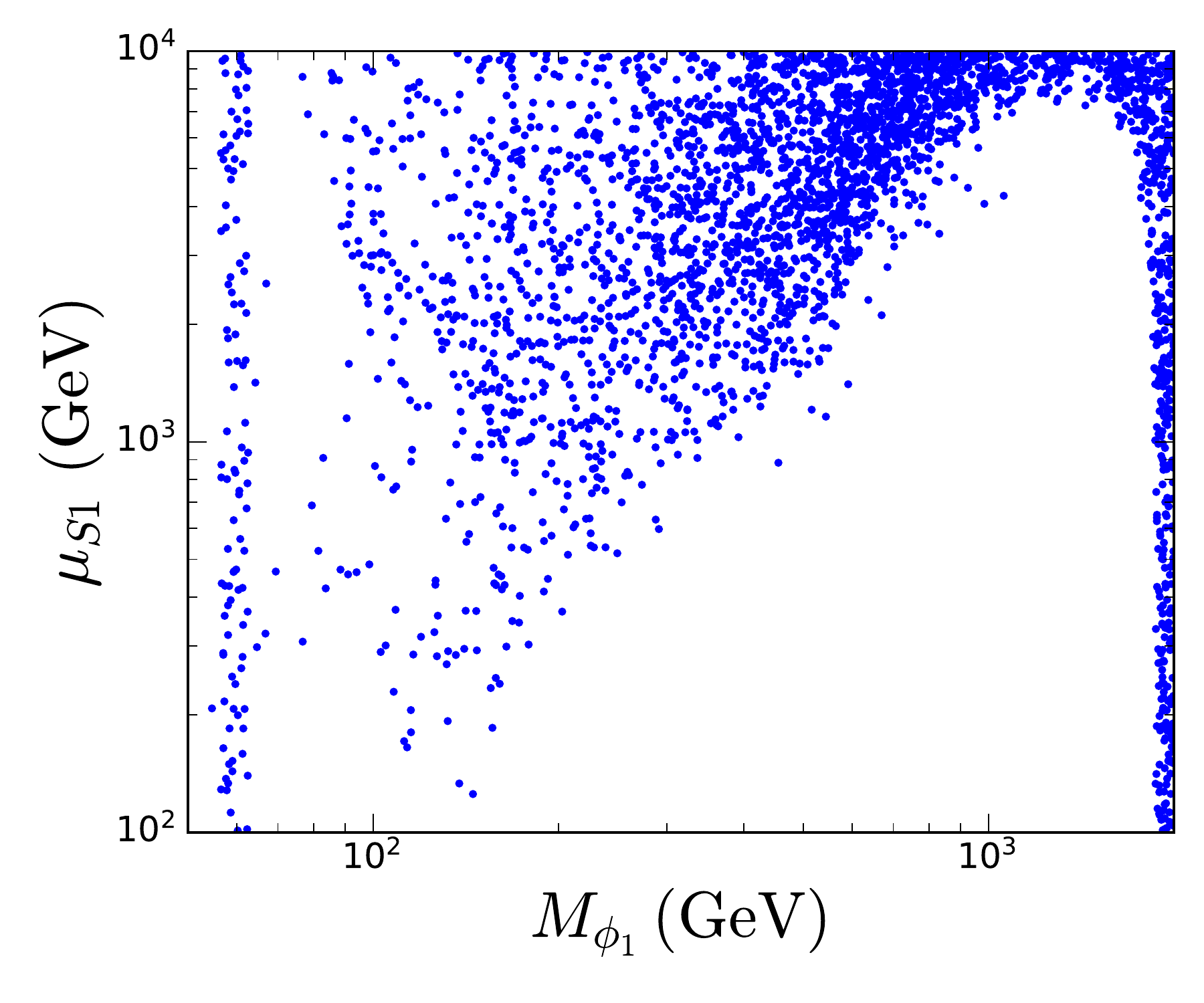}\\
\includegraphics[scale=0.44]{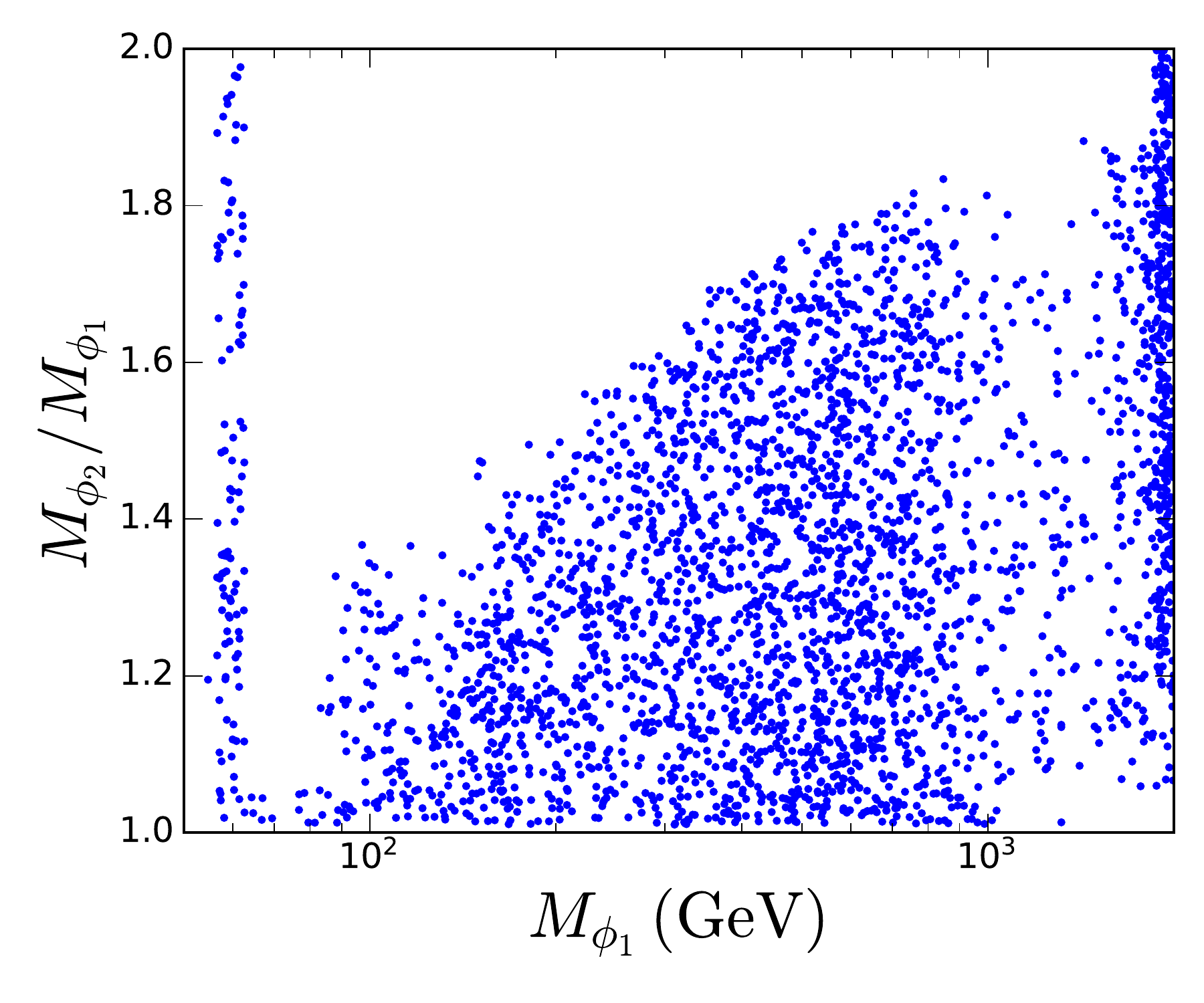}
\includegraphics[scale=0.44]{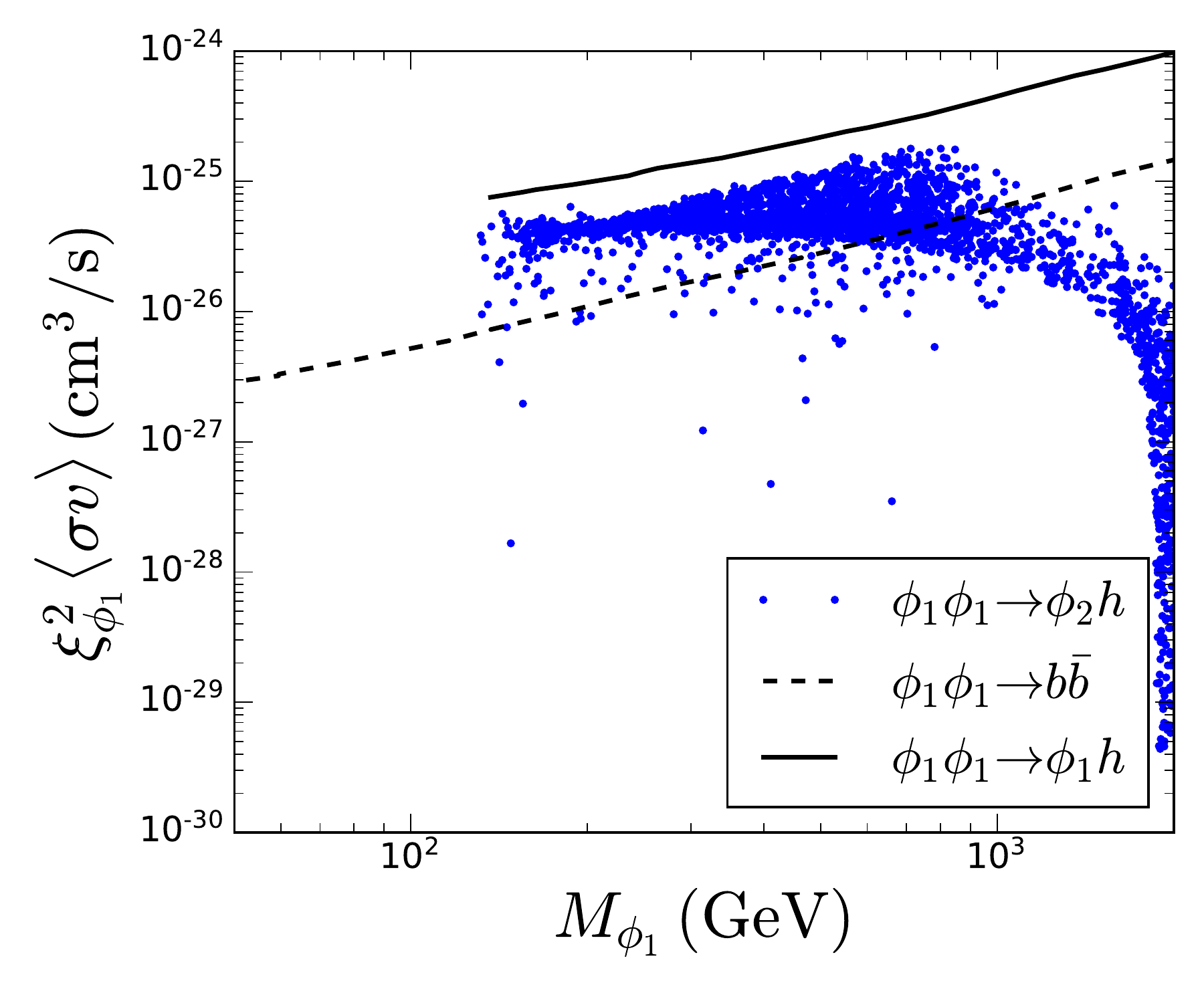}\\
\includegraphics[scale=0.44]{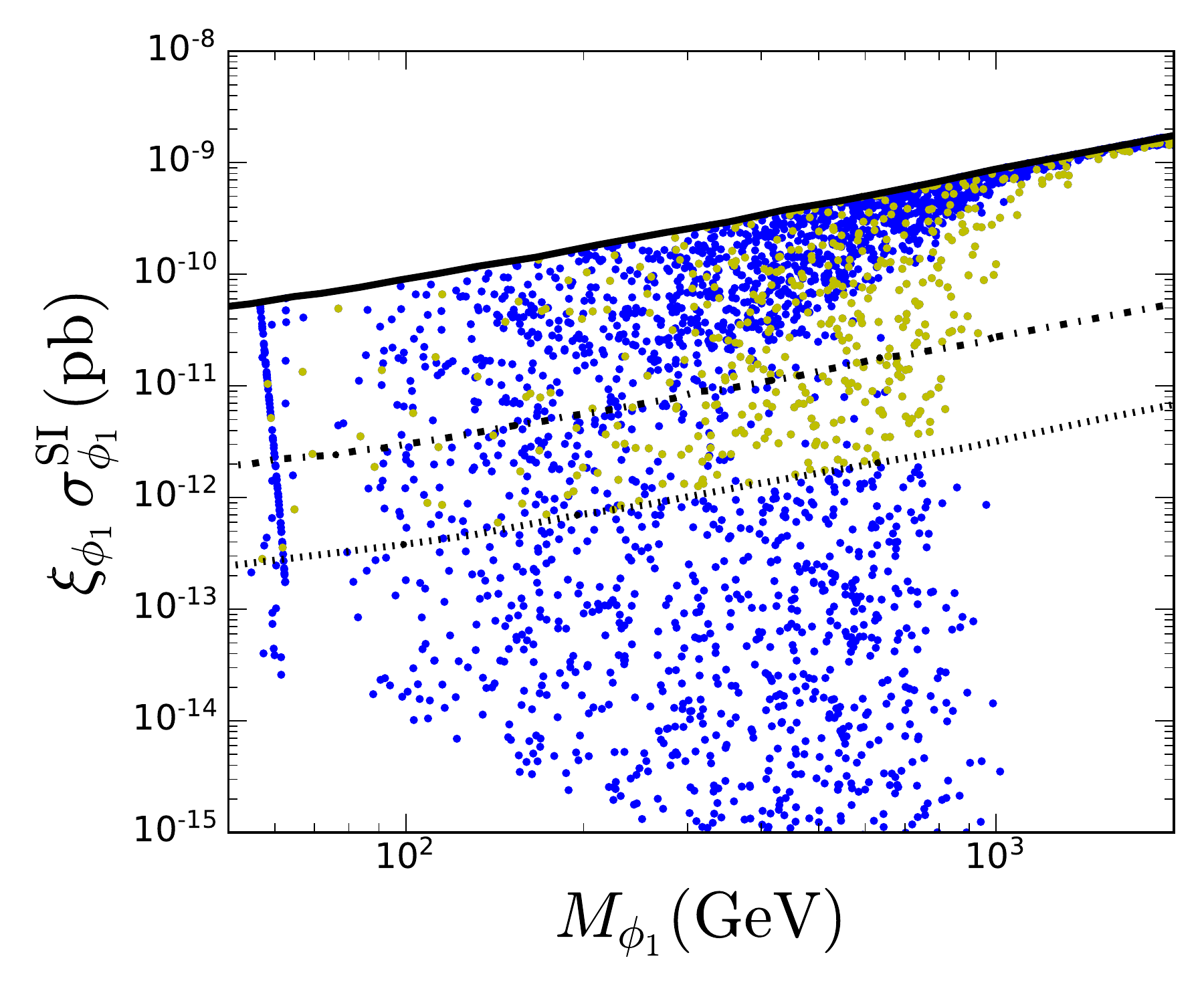}
\includegraphics[scale=0.44]{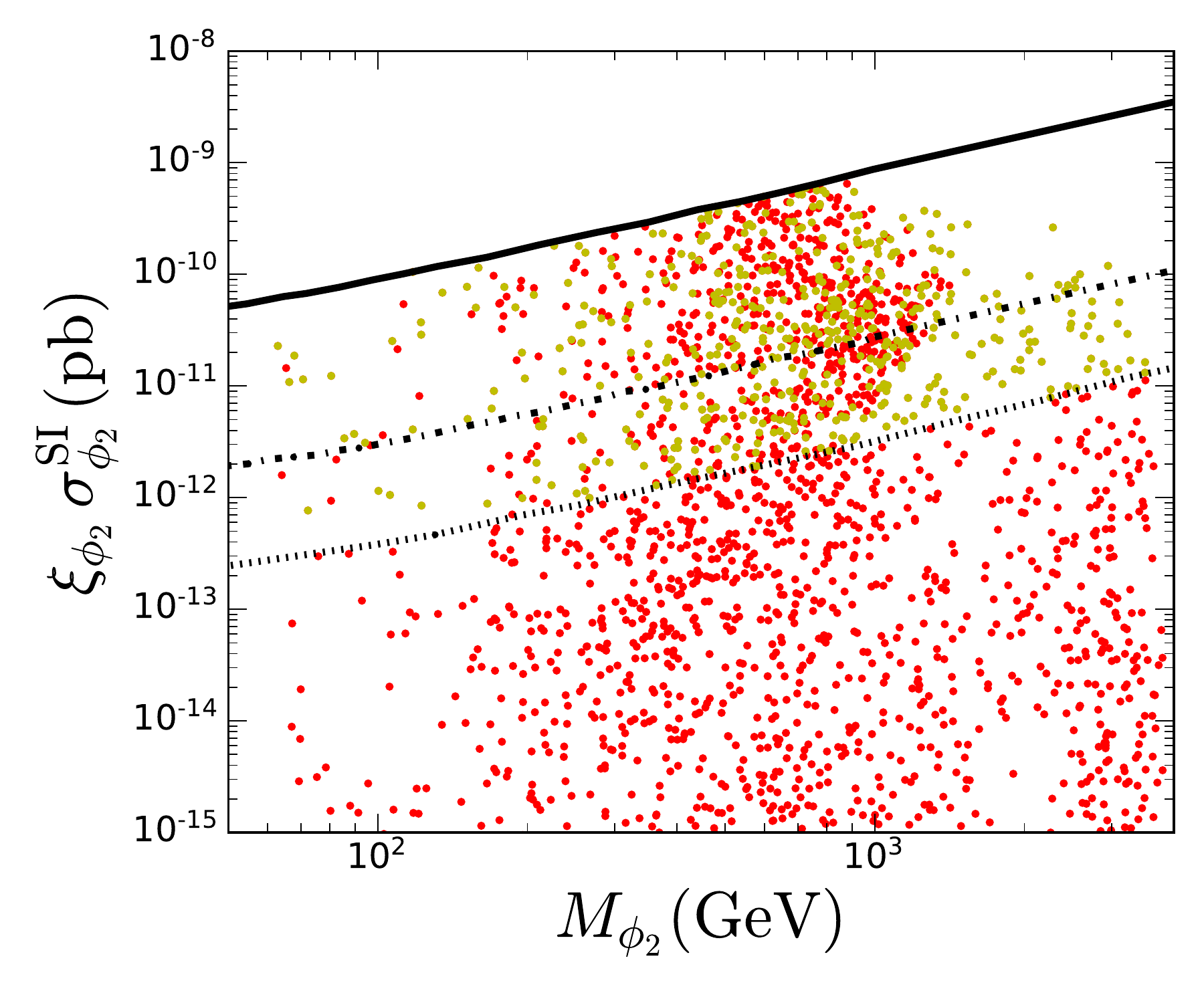}
\caption{A sample of viable points of the $Z_4$ model with $M_{\phi_1}<M_{\phi_2}$, projected along different dimensions. The top panels show scatter plots of $M_{\phi_1}$ versus $\Omega_{\phi_1}/\Omega_{DM}$ (left) and versus $\mu_{S1}$ (right). In the center panels, the ratio of  dark matter masses (left) and the most relevant indirect detection signal (right) are illustrated. The direct detection prospects are shown in the bottom panels for $\phi_1$ (left) and $\phi_2$ (right). } 
\label{fig:Z4scan-P1}
\end{figure}

\begin{figure}[t]
\centering
\includegraphics[scale=0.44]{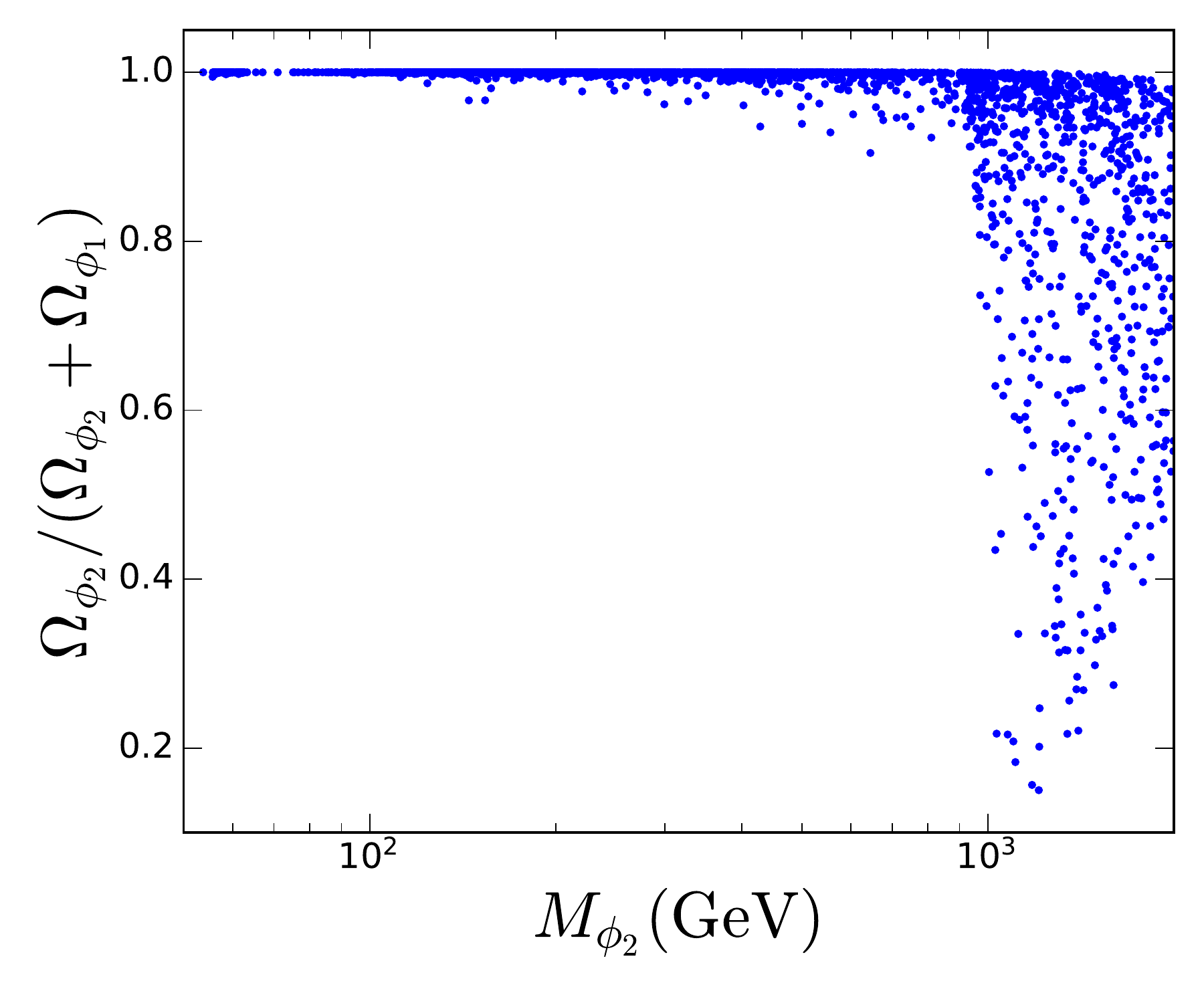}
\includegraphics[scale=0.44]{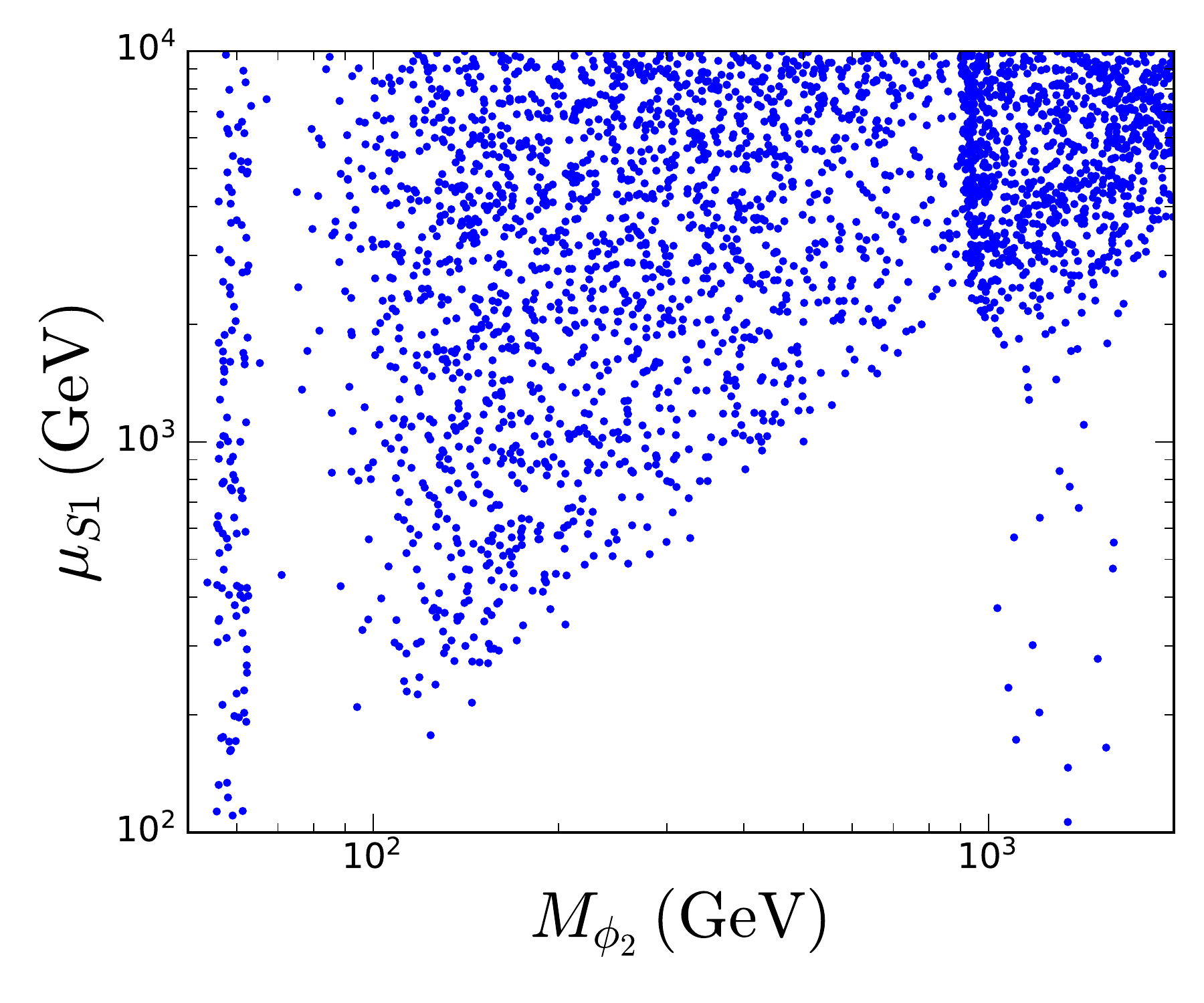}\\
\includegraphics[scale=0.44]{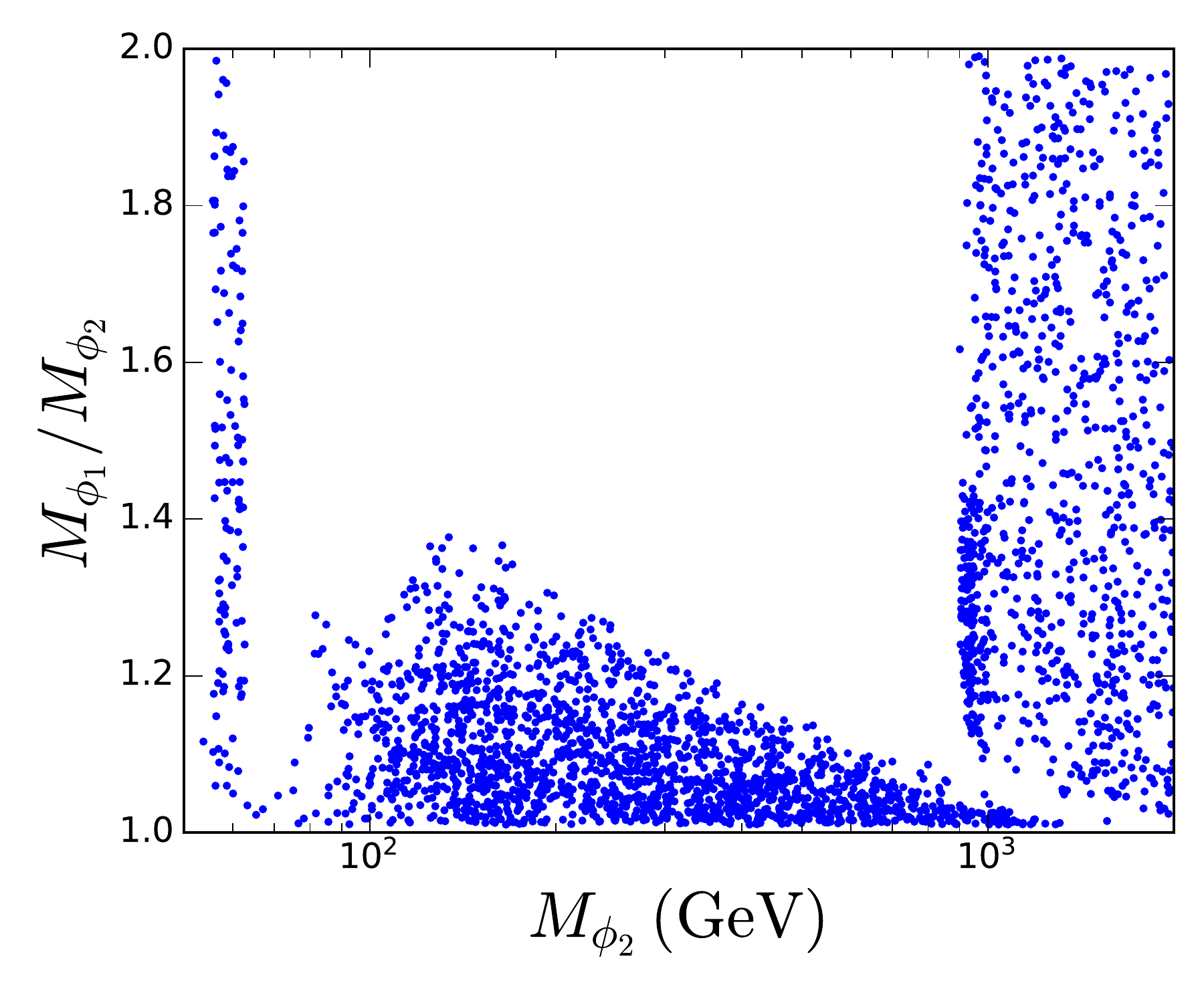}
\includegraphics[scale=0.44]{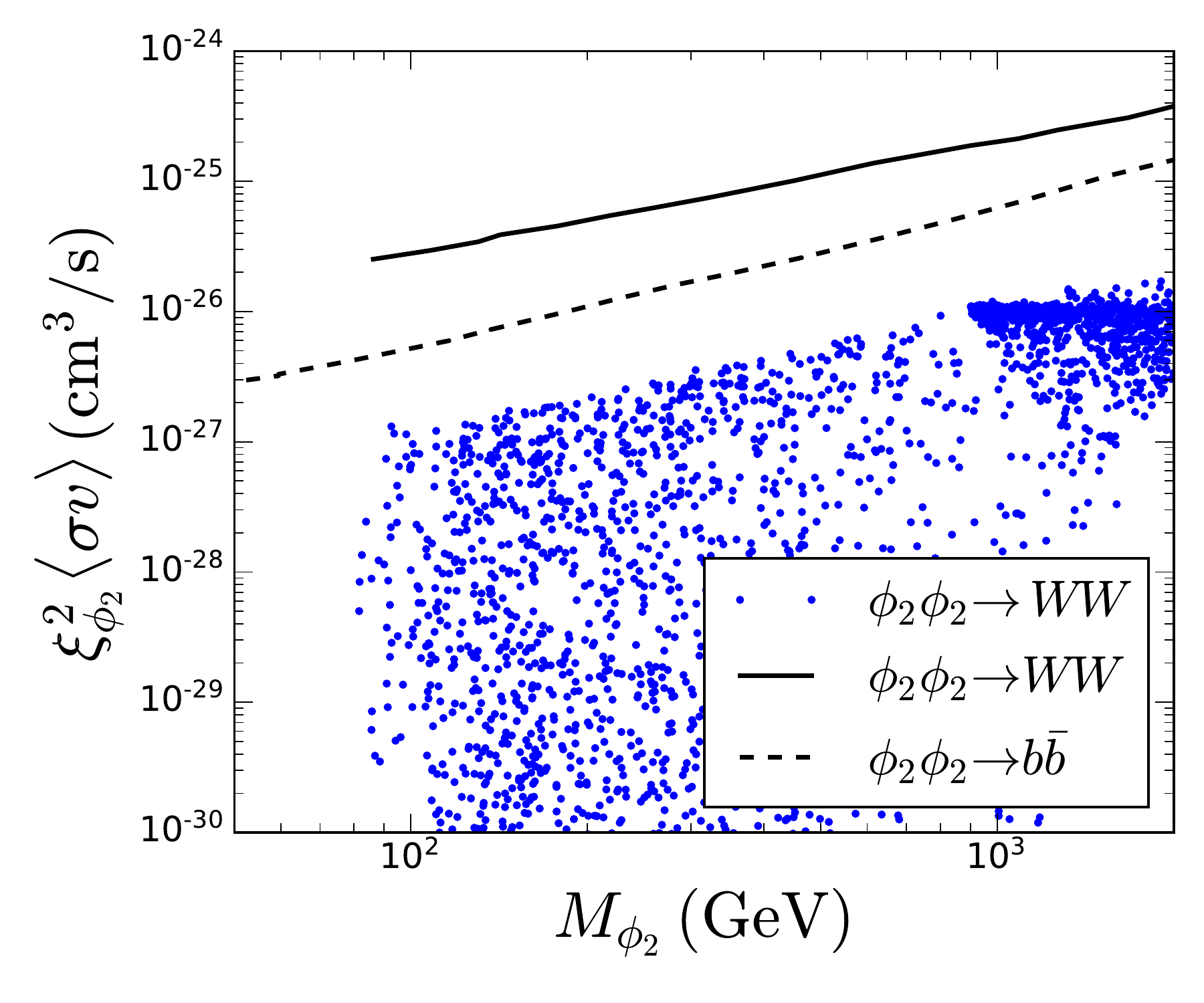}\\
\includegraphics[scale=0.44]{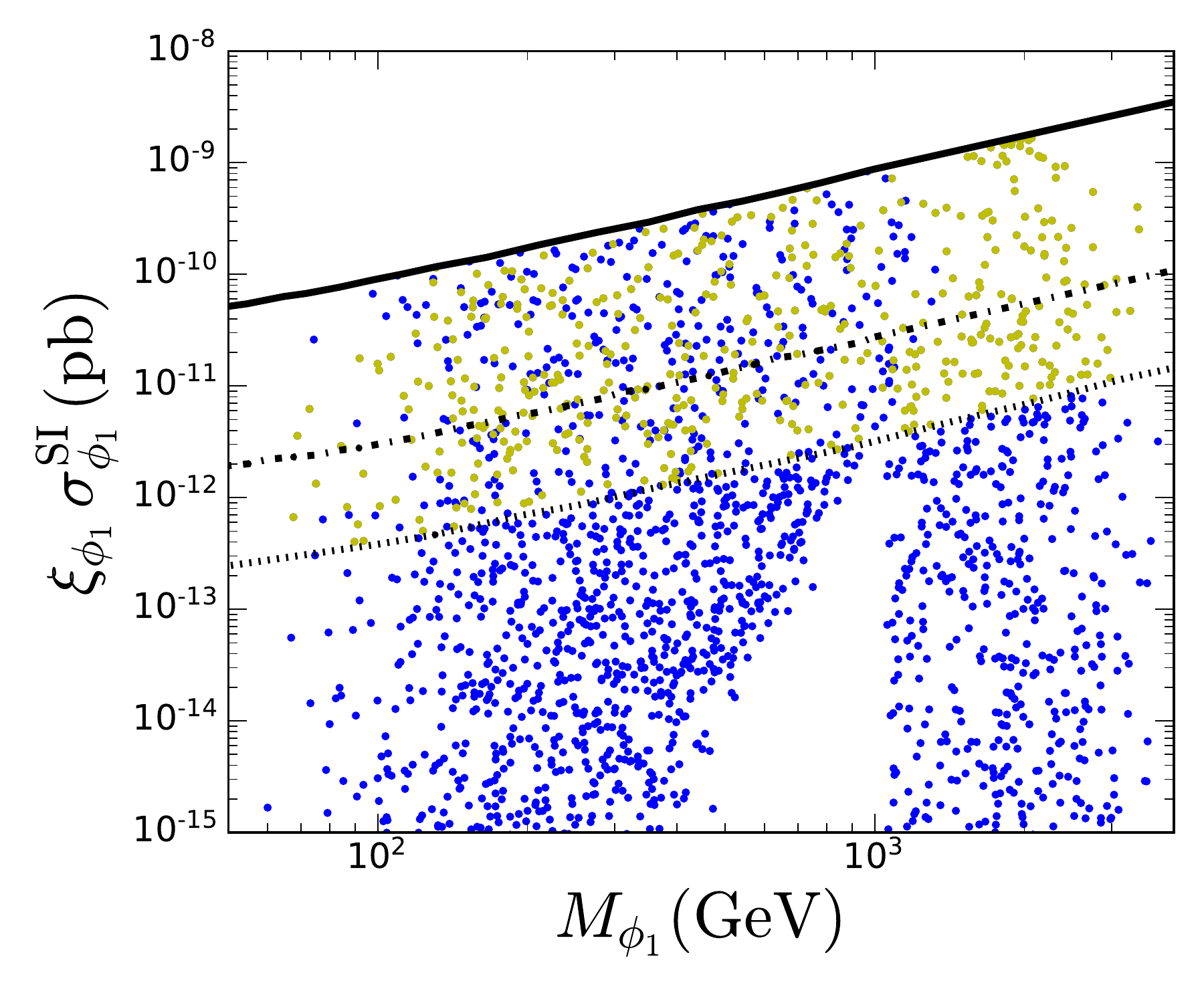}
\includegraphics[scale=0.44]{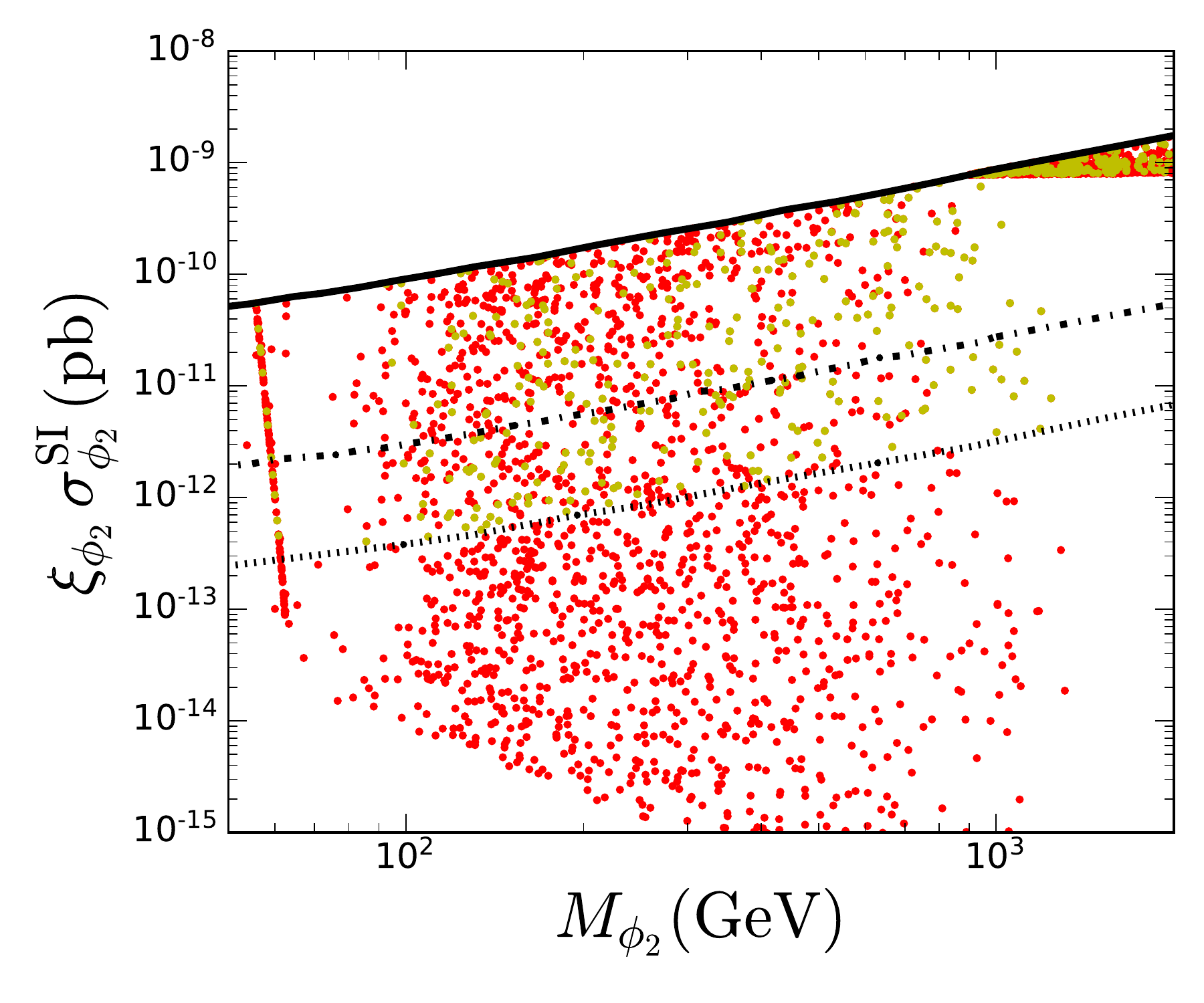}\\
\caption{A sample of viable points of the $Z_4$ model with $M_{\phi_2}<M_{\phi_1}$, projected along different dimensions. The top panels show scatter plots of $M_{\phi_2}$ versus $\Omega_{\phi_2}/\Omega_{DM}$ (left) and versus $\mu_{S1}$ (right). In the center panels, the ratio of  dark matter masses (left) and the most relevant indirect detection signal (right) are illustrated. The direct detection prospects are shown in the bottom panels for $\phi_1$ (left) and $\phi_2$ (right). 
}
\label{fig:Z4scan-S1}
\end{figure}

In this model the $\mathcal{V}_2$ potential contains the trilinear interaction $\phi_1^2\phi_2$ which induces  semi-annihilation processes that may help decrease the number density of dark matter particles. The associated coupling is chosen as 
\begin{align}
    &100\,{\rm GeV}\leq \mu_{S1} \leq 10\,{\rm TeV}.
\end{align}
The remaining free parameter of the model, $\lambda_{41}$, is varied in the interval\footnote{Notice that $\lambda_{51}$ plays no role in the current analysis.} 
\begin{align}
    &10^{-4}\leq |\lambda_{412}|\leq 1.
\end{align}
Both couplings were sampled using a log-uniform distribution in the given range. It is convenient to separately investigate the two possible mass hierarchies for the dark matter particles in this scenario --$M_{\phi_1}<M_{\phi_2}$ and $M_{\phi_2}<M_{\phi_1}$. The resulting viable parameter space is shown in figures \ref{fig:Z4scan-P1} and \ref{fig:Z4scan-S1}, respectively. The most important conclusion that can be drawn from these figures is that, in contrast with the singlet scalar model, it is indeed possible to satisfy current bounds with dark matter masses below the TeV scale. In fact, the entire range of dark matter masses explored in the scan turns out to be viable.  

From the top-left panel of  figure \ref{fig:Z4scan-P1} we see that $\phi_1$  gives the dominant contribution to the dark matter density for most of the viable points in our scan, usually accounting for more than $90\%$ of it.  In fact, in multiple cases $\Omega_{\phi_2}$ turns out to be several orders of magnitude smaller than $\Omega_{\phi_1}$. This hierarchy is a consequence of the new $Z_4$ interactions, which tend to suppress the relic density of the heavier particle more than that of the lighter one --because the former can annihilate into the latter.

The fact that  semi-annihilation processes (see Fig.~\ref{fig:semi2}) are essential to obtain the correct relic density while satifying direct detection bounds is illustrated in the top-right panel, which shows the trilinear coupling versus $M_{\phi_1}$.   Notice that the minimum value of $\mu_{S1}$ found in our sample increases with $M_{\phi_1}$ up to about $1$ TeV, when it reaches the maximum value allowed in the scan ($10$ TeV). At $M_{\phi_1}\sim 2~$TeV or higher, the trilinear couplings can be small because the standard Higgs portal becomes consistent with current data.  The center-left panel displays the ratio $M_{\phi_2}/M_{\phi_1}$, whose range of variation tends to increase with $M_{\phi_1}$.  This figure  demonstrates that, in this scenario,  the masses of the dark matter particles are not required to be degenerate. 

Regarding indirect detection, the dominant annihilation channel in our sample  turns out to be the semi-annihilation process $\phi_1+\phi_1\to \phi_2+h$. Due to the $\Omega_{\phi_2}$ suppression, all the $\phi_2$ annihilation channels feature instead much lower rates. The center-right panel displays the $\phi_1+\phi_1\to \phi_2+h$ rate  versus $M_{\phi_1}$ for our sample of viable points. Since up to date there is no reported experimental limit on such a process, we show, for comparison purposes, the current limit on the related process $\phi_1+\phi_1\to \phi_1+h$ (solid line) \cite{Queiroz:2019acr}, which should be slightly stronger than the one on  $\phi_1+\phi_1\to \phi_2+h$. From the figure we conclude that none of the viable points found in our scan is excluded by the current limit.  The dashed line corresponds instead to the projected sensitivity, assuming  45 dSphs and 15 years of observation\cite{Charles:2016pgz}, for the process $\phi_1+\phi_1\to b+\bar b$, which yields a higher $\gamma$-ray flux than $\phi_1+\phi_1\to \phi_2+h$.  It is unclear, therefore, whether future observations will be able to set constraints on the viable points of this scenario.




In the bottom panels, the direct detection rates for both dark matter particles are compared against the experimental  results.   The solid line shows the current limit from XENON1T, which is necessarily satisfied due to our selection procedure,  while the dashed and dotted lines correspond to the expected sensitivities of   LZ~\cite{Akerib:2018lyp} and DARWIN \cite{Aalbers:2016jon} respectively. In this figures, the yellow points denote the viable models for which \emph{both} dark matter particles are expected to yield signals in future direct detection experiments. From these two panels we conclude that either dark matter particle may be observed in future direct detection experiments, and that in several cases  both dark matter particles might be observed.     

\begin{figure}[t]
\centering
\includegraphics[scale=0.44]{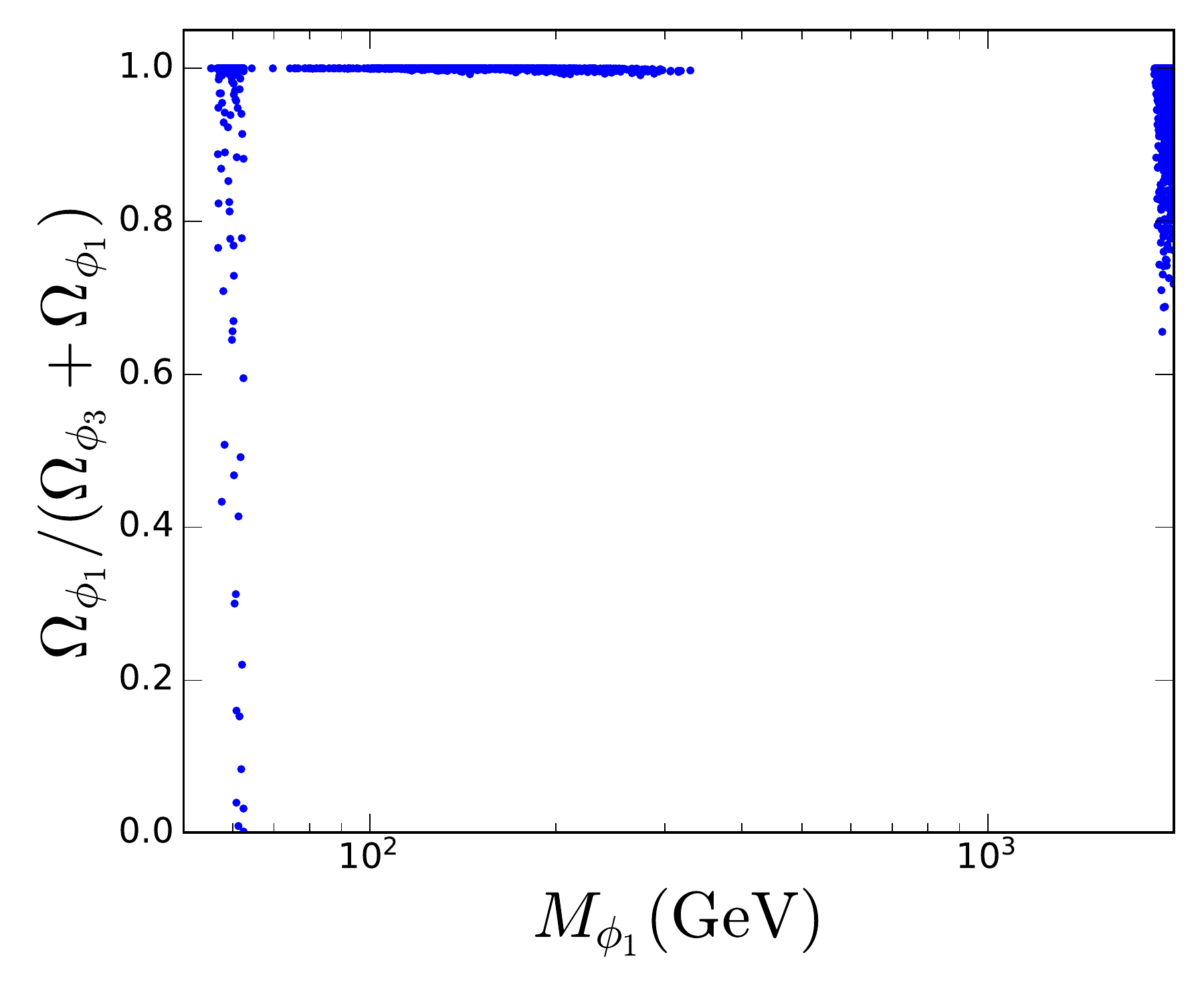}
\includegraphics[scale=0.44]{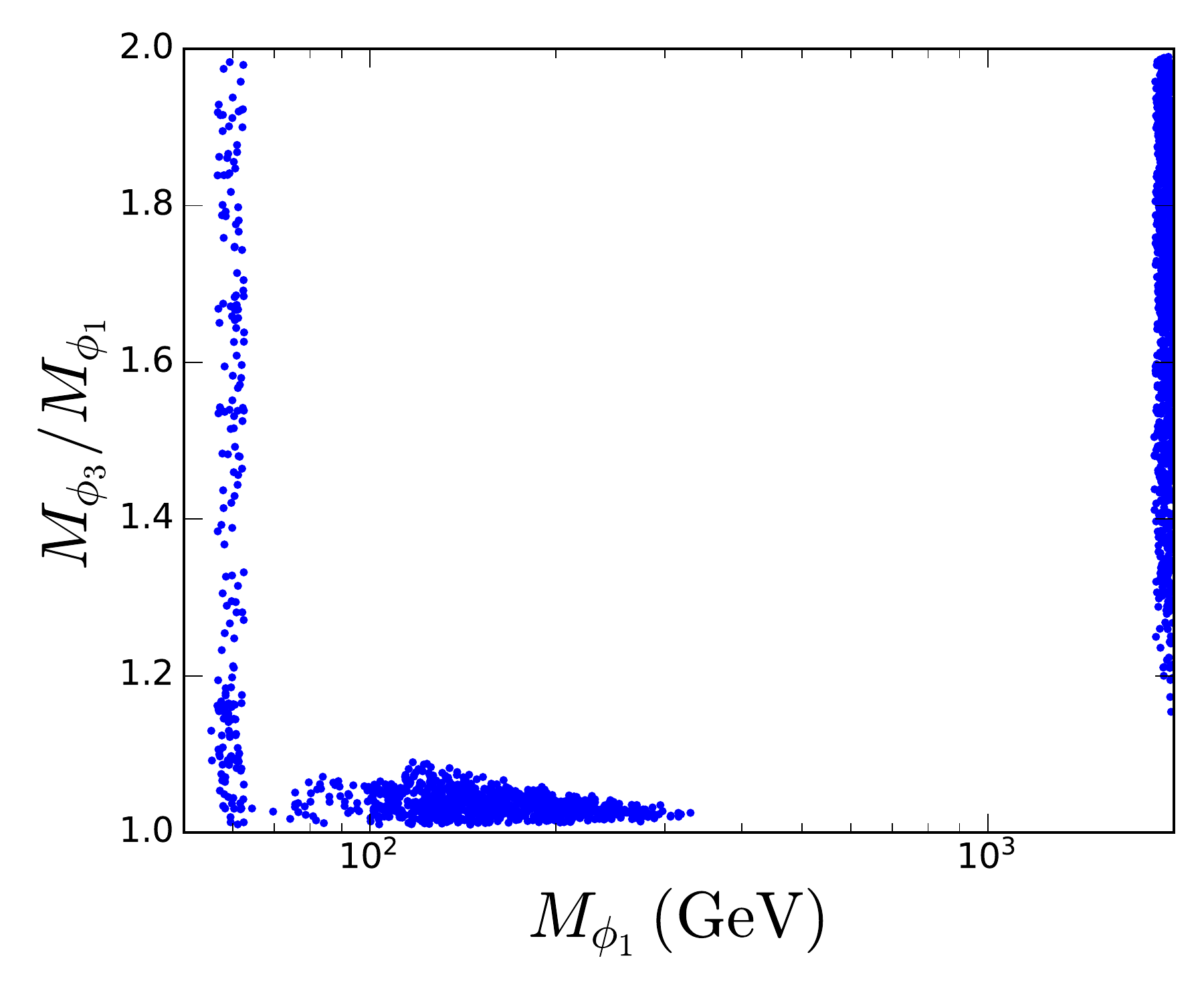}\\
\includegraphics[scale=0.44]{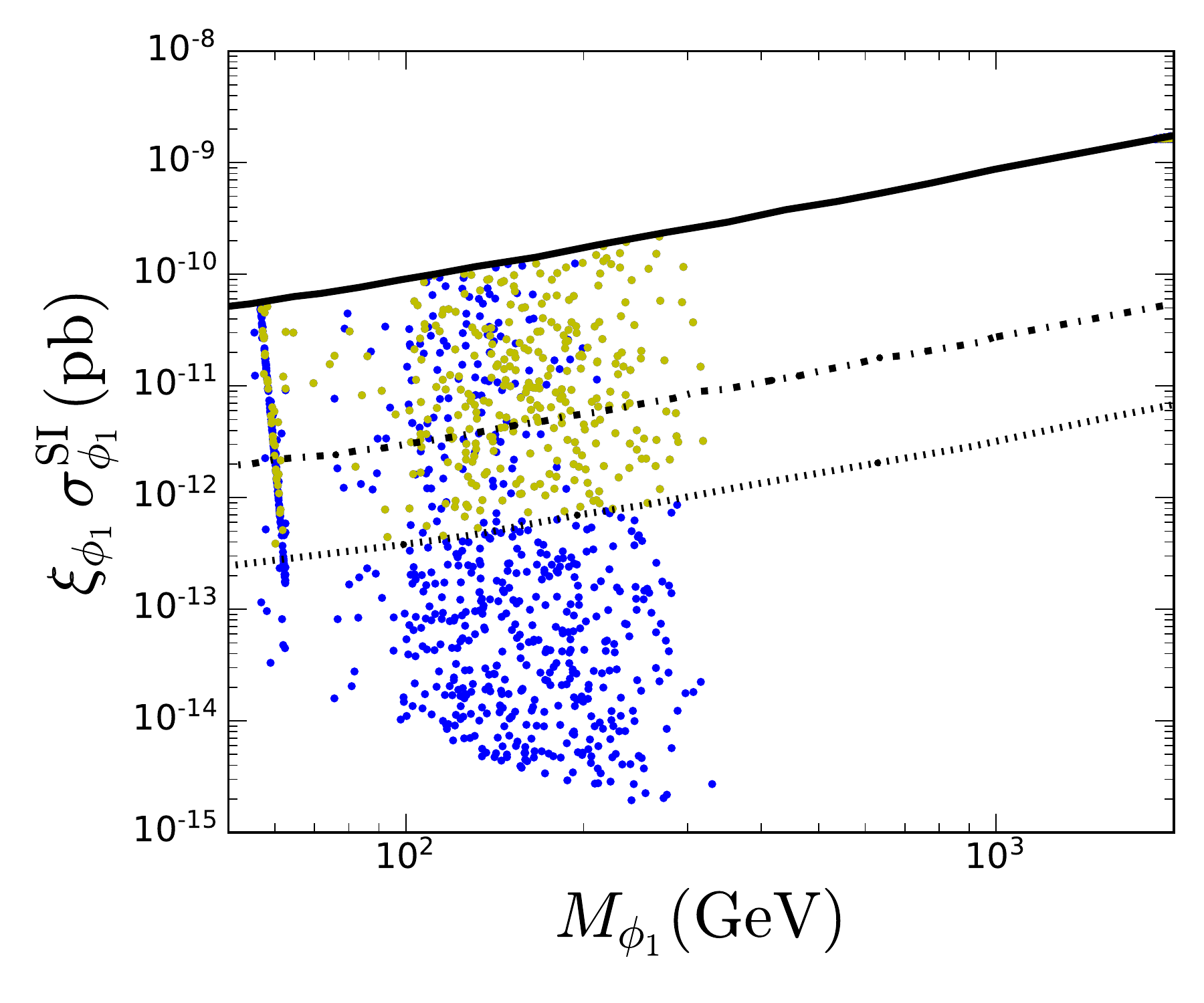}
\includegraphics[scale=0.44]{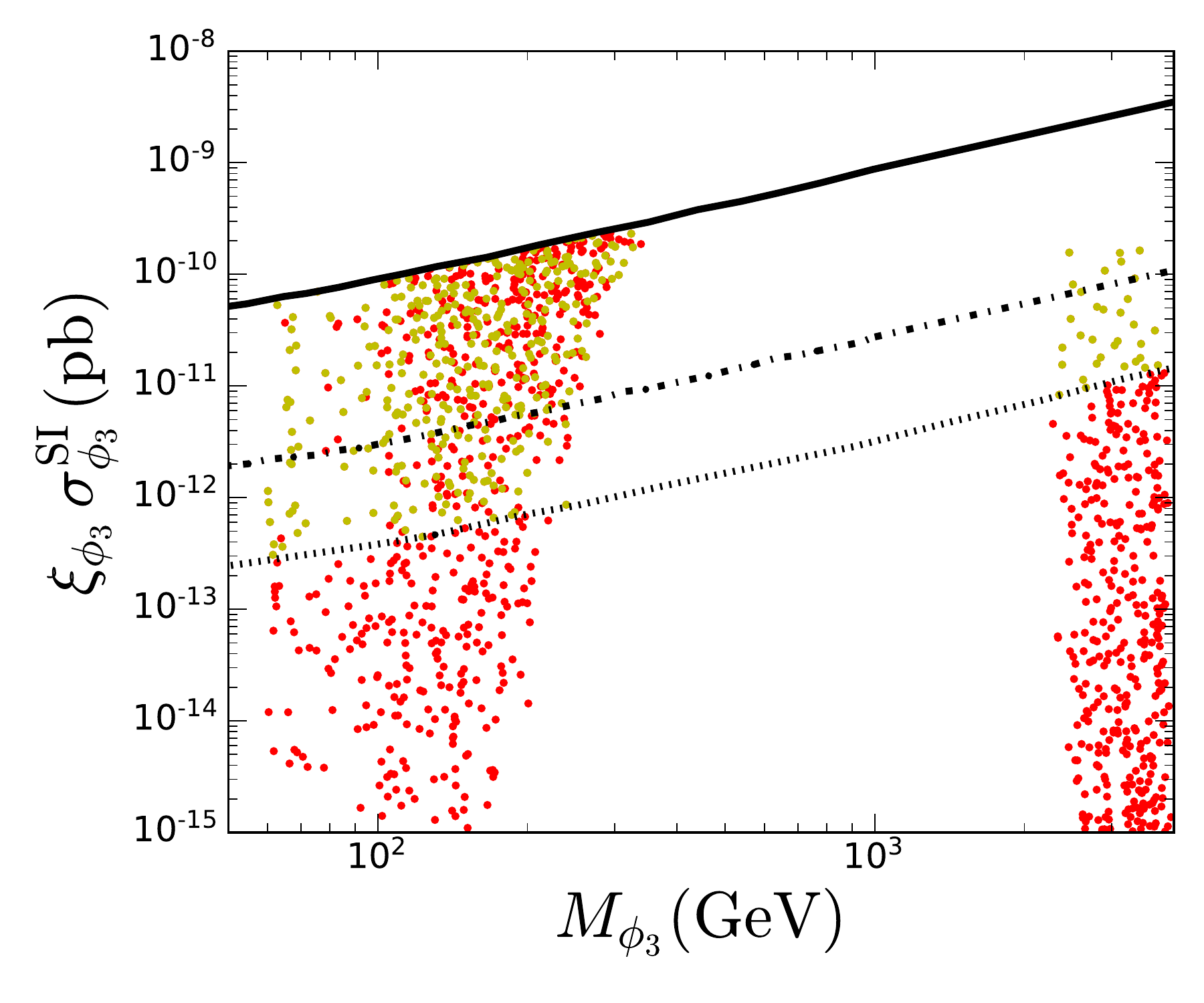}
\caption{A sample of viable points of the $Z_6$ model with $(\phi_1,\phi_3)$ and $M_{\phi_1}<M_{\phi_3}$, projected along different dimensions. The top panels show scatter plots of $M_{\phi_1}$ versus $\Omega_{\phi_1}/\Omega_{DM}$ (left) and versus the ratio of  dark matter masses (right).  The direct detection prospects are shown in the bottom panels for $\phi_1$ (left) and $\phi_3$ (right).}
\label{fig:scanZ613P1}
\end{figure}

The results for the case where $M_{\phi_2}<M_{\phi_1}$ (see Fig. \ref{fig:Z4scan-S1}) are qualitatively similar to those for $M_{\phi_1}<M_{\phi_2}$. The main difference is that in this case a mild degeneracy is required between the dark matter particles. In fact, notice from the center-left panel that $M_{\phi_2}/M_{\phi_1}$ does not exceed $1.4$ within our sample of viable points. It is worth mentioning that despite the fact that for values of $M_{\phi_2}$ above $950$ GeV the $Z_4$ interactions are not compulsory to achieve a depletion on $\Omega_{\phi_2}$, they are still necessary to allow for efficient $\phi_1$ self-annihilations. 
As can be seen from the center-right panel, the indirect detection prospects are not encouraging in this case. It shows the rate for the dominant annihilation channel in our sample, which  happens to be $\phi_2+\phi_2^\dagger\to W^+W^-$, and compares it against  the corresponding current limit reported by the Fermi collaboration \cite{Ackermann:2015zua} (solid line). All the viable points lie well below the current bound. Even future Fermi data (dashed line) will be unable to exclude viable points.  Future direct detection experiments, on the other hand,  will set significant constraints on this scenario (see bottom panels) and constitute the most promising way to test it.

Finally, our scans indicate that, independently of the mass hierarchy,  the $\lambda_{412}$ interaction does not play an essential role in setting the dark matter abundances. It is instead the trilinear couplings, via the semi-annihilation processes they induce, that modify the relic density and allow to have dark matter masses below a TeV.

\begin{figure}[t]
\centering
\includegraphics[scale=0.44]{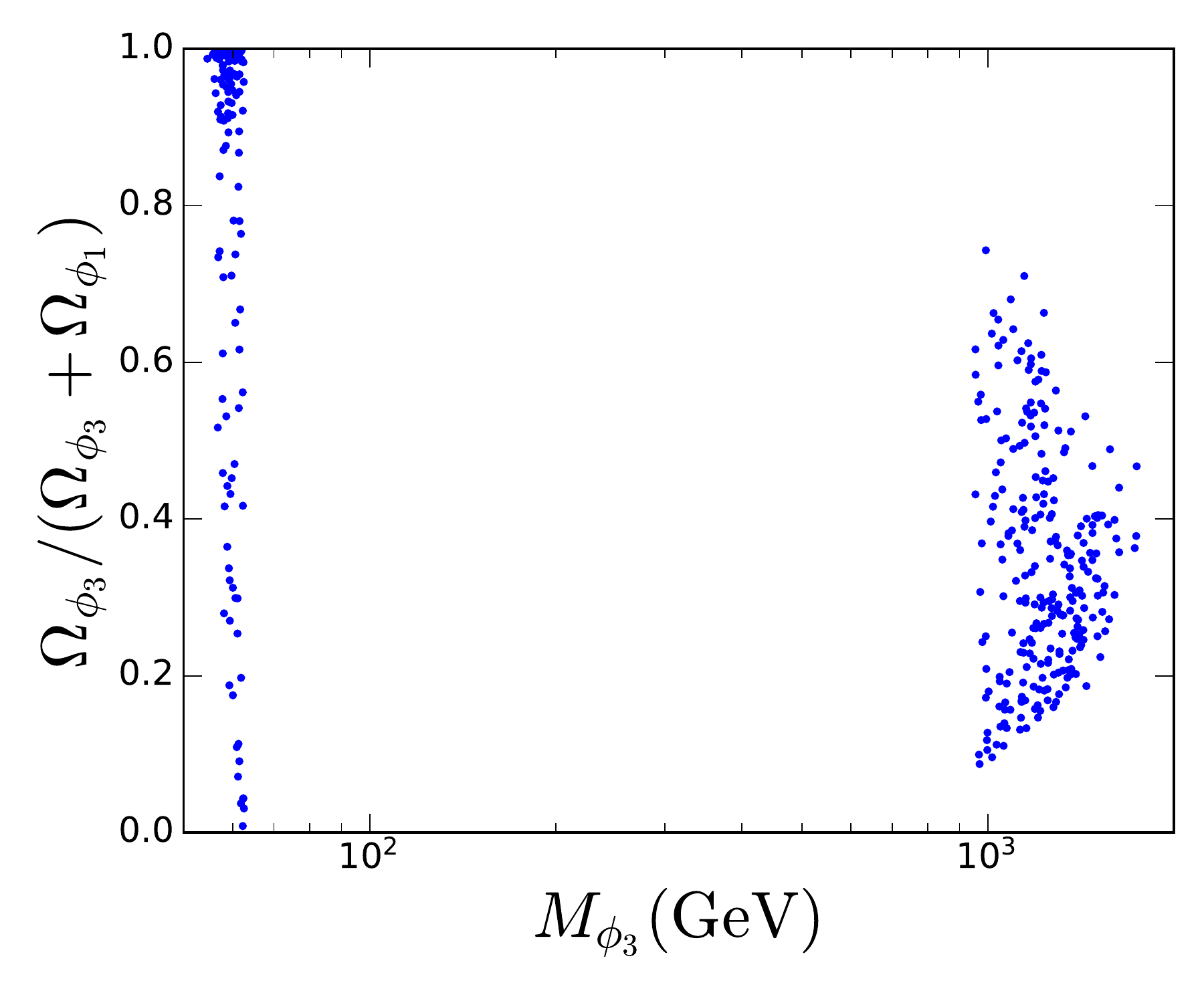}
\includegraphics[scale=0.44]{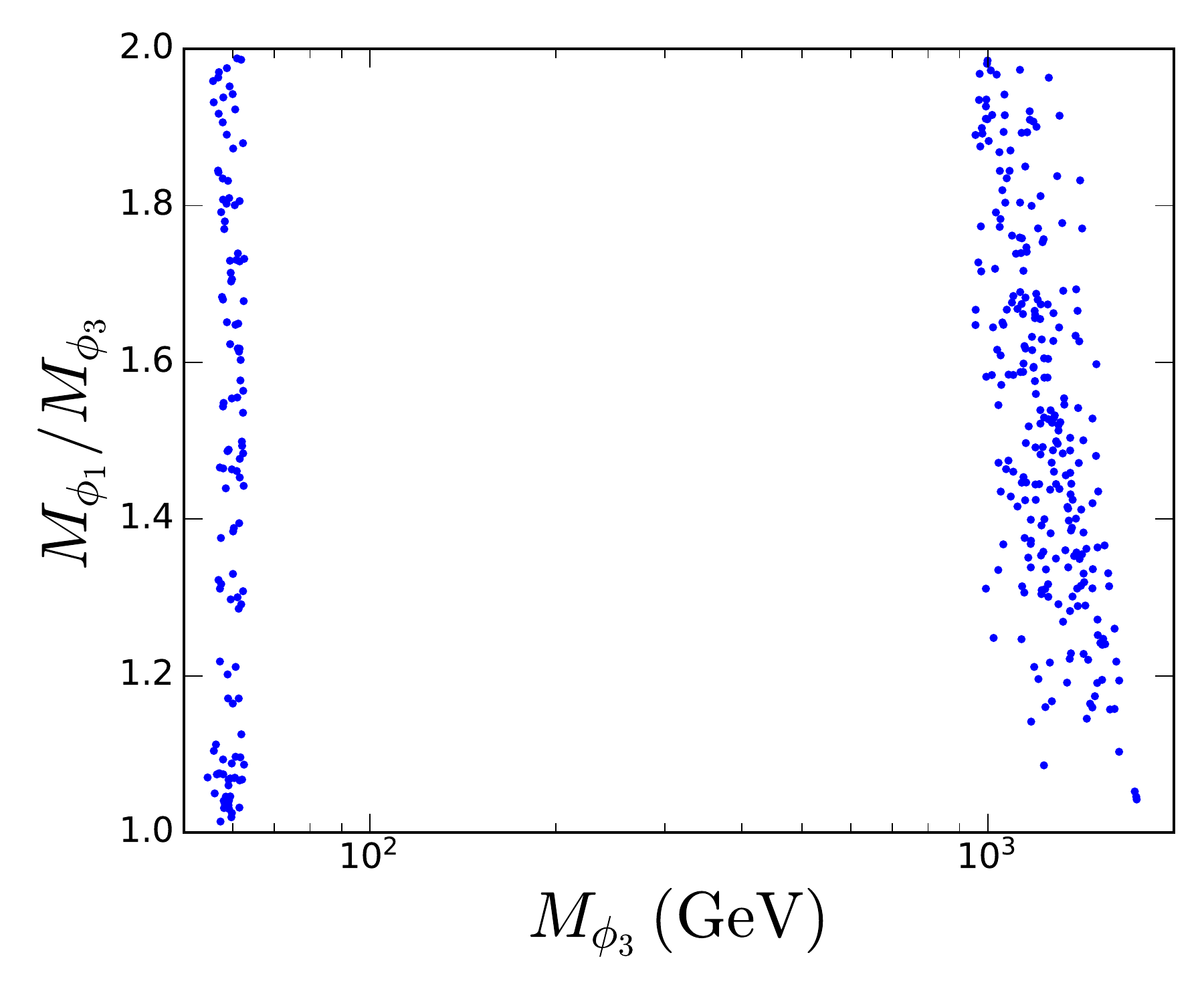}\\
\includegraphics[scale=0.44]{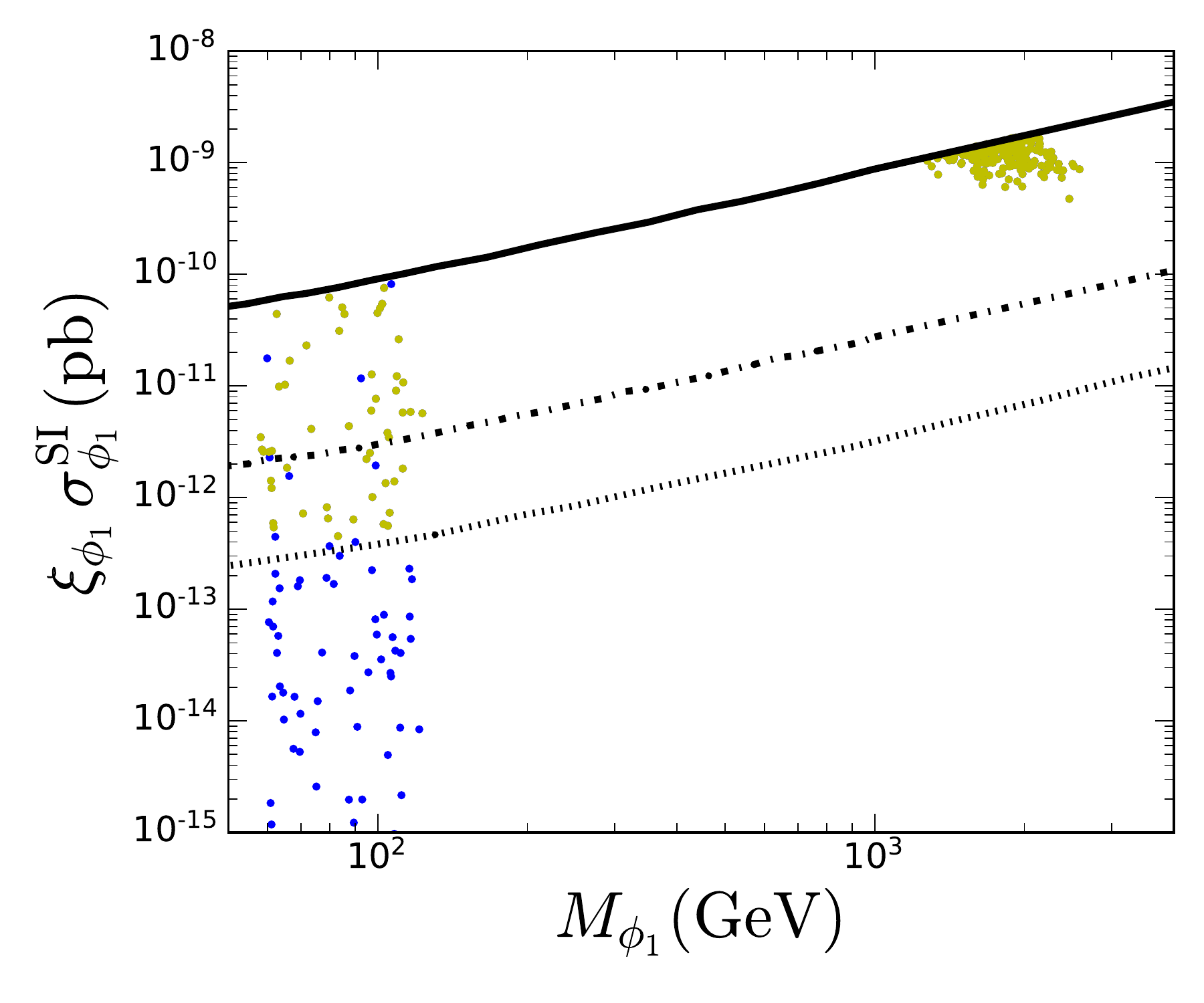}
\includegraphics[scale=0.44]{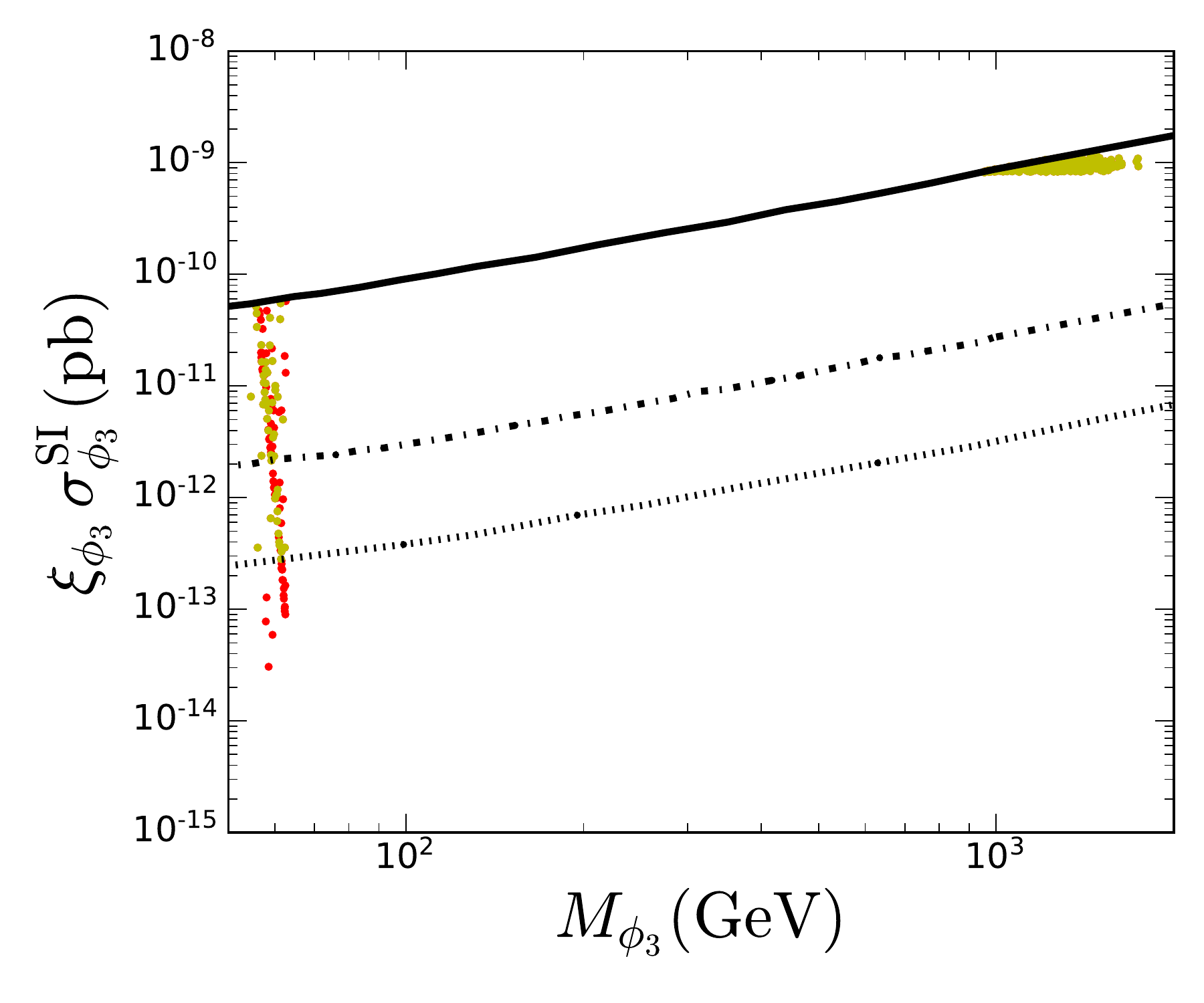}
\caption{A sample of viable points of the $Z_6$ model with $(\phi_1,\phi_3)$ and $M_{\phi_3}<M_{\phi_1}$, projected along different dimensions. The top panels show scatter plots of $M_{\phi_1}$ versus $\Omega_{\phi_1}/\Omega_{DM}$ (left) and versus the ratio of  dark matter masses (right).  The direct detection prospects are shown in the bottom panels for $\phi_1$ (left) and $\phi_3$ (right).}
\label{fig:scanZ613S1}
\end{figure}

\subsection{$Z_6(13)$ model}
The unique new interaction associated to this model is $\lambda'_{41}$ (see Eq.~(\ref{eq:V2Z613})), which only generates dark matter conversion processes of the type $\phi_1\leftrightarrow \phi_{3}$. In our scans $\lambda_{413}$ and  $\lambda'_{41}$ are randomly sampled using a log-uniform distribution in the range
\begin{align}
    &10^{-4}\leq |\lambda_{413}|,\, |\lambda'_{41}|\leq 1.
\end{align}
The results from the scan are displayed in figures \ref{fig:scanZ613P1} and \ref{fig:scanZ613S1} for the $M_{\phi_1}<M_{\phi_3}$ and $M_{\phi_3}<M_{\phi_1}$,  respectively. 
In the former case, the main effect of the quartic interactions is to open up a new  viable region with  $70~\mathrm{GeV}\lesssim M_{\phi_1}\lesssim 300~\mathrm{GeV}$ --the region between $300$ and $1850$ GeV features no viable points in our sample.  And  it is $\phi_1$, the lighter dark matter particle, that accounts for most of the relic density in the new viable region --see top right panel.  In this case, the depletion of $\Omega_{\phi_1}$ is rather inefficient because it proceeds via  $\lambda'_{41}$-induced dark matter conversion processes, which are only possible when the dark matter particles are mass degenerate (so that the kinematic suppression is not that strong).  In fact, notice from the top-right panel that the degree of degeneracy is always below $10\%$ and decreases with $M_{\phi_1}$.  As was the case in the $Z_4$ model, the $\lambda_{413}$ interaction plays no role here.  Regarding detection prospects, the bottom panels show that a significant fraction of the viable points in our scan lies within the expected sensitivity of future experiments. Due to the mass degeneracy, however, it would be extremely difficult if not altogether impossible to disentangle the possible signals from the two dark matter particles. 

\begin{figure}
\centering
\includegraphics[scale=0.44]{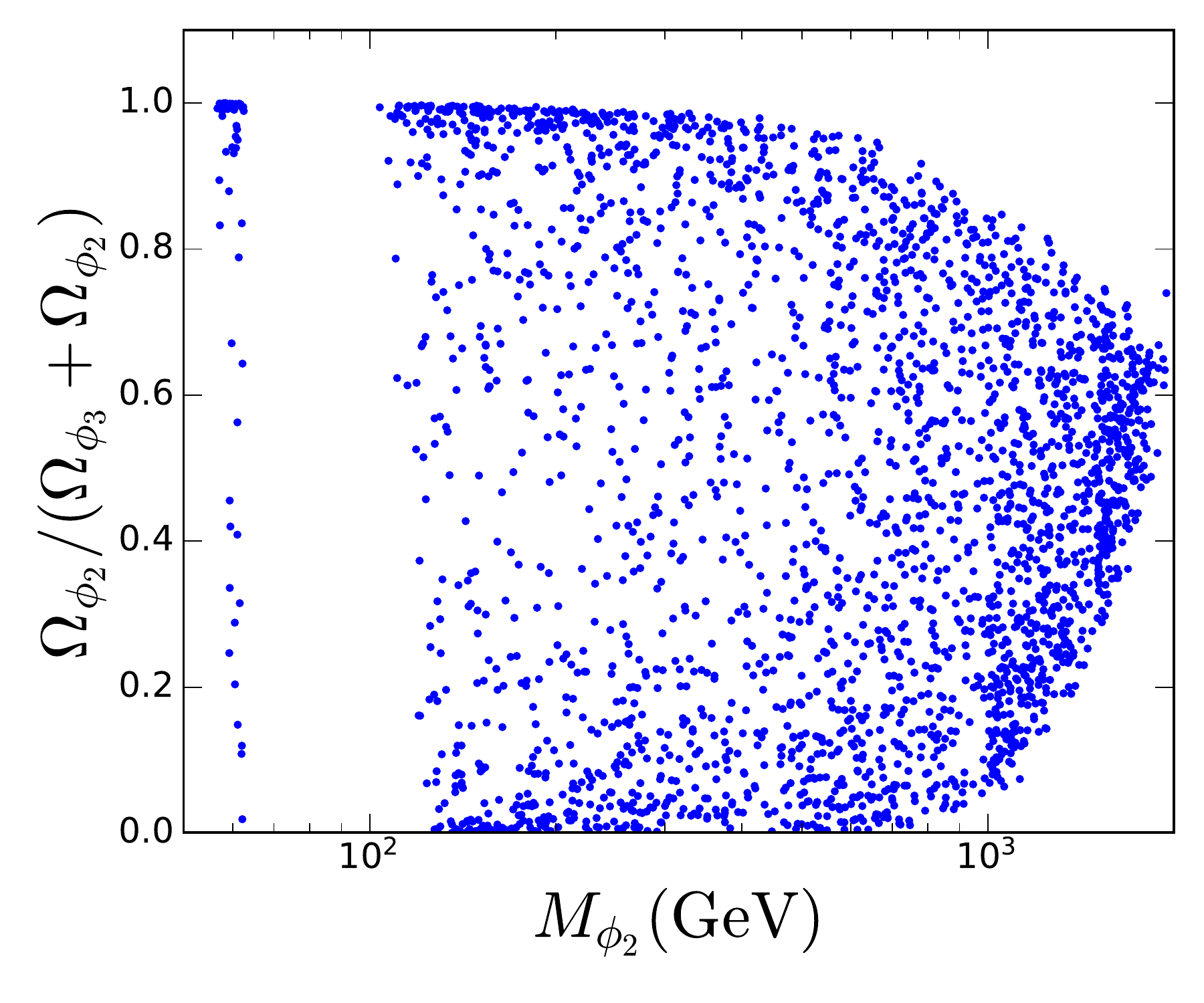}
\includegraphics[scale=0.44]{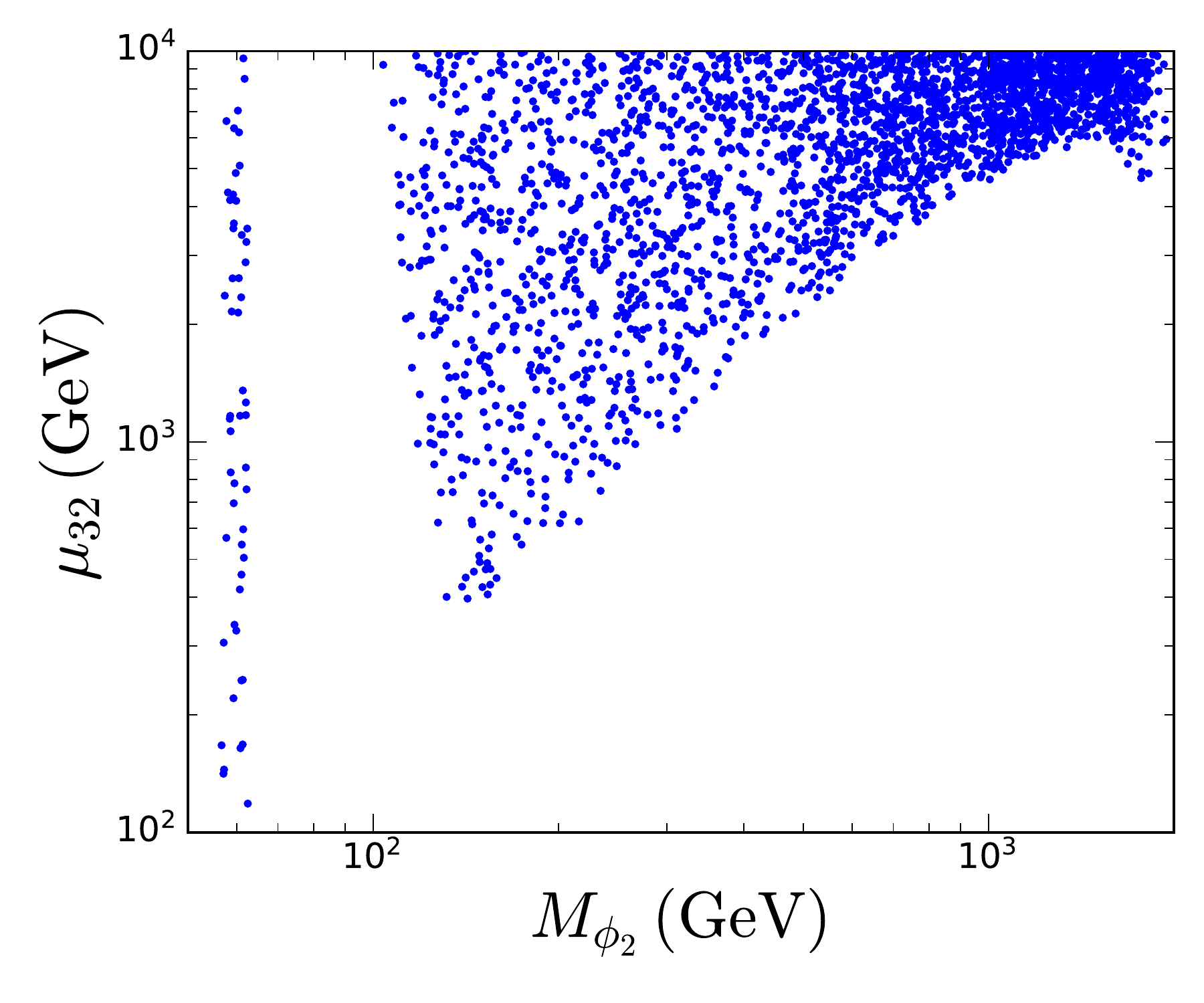}\\
\includegraphics[scale=0.44]{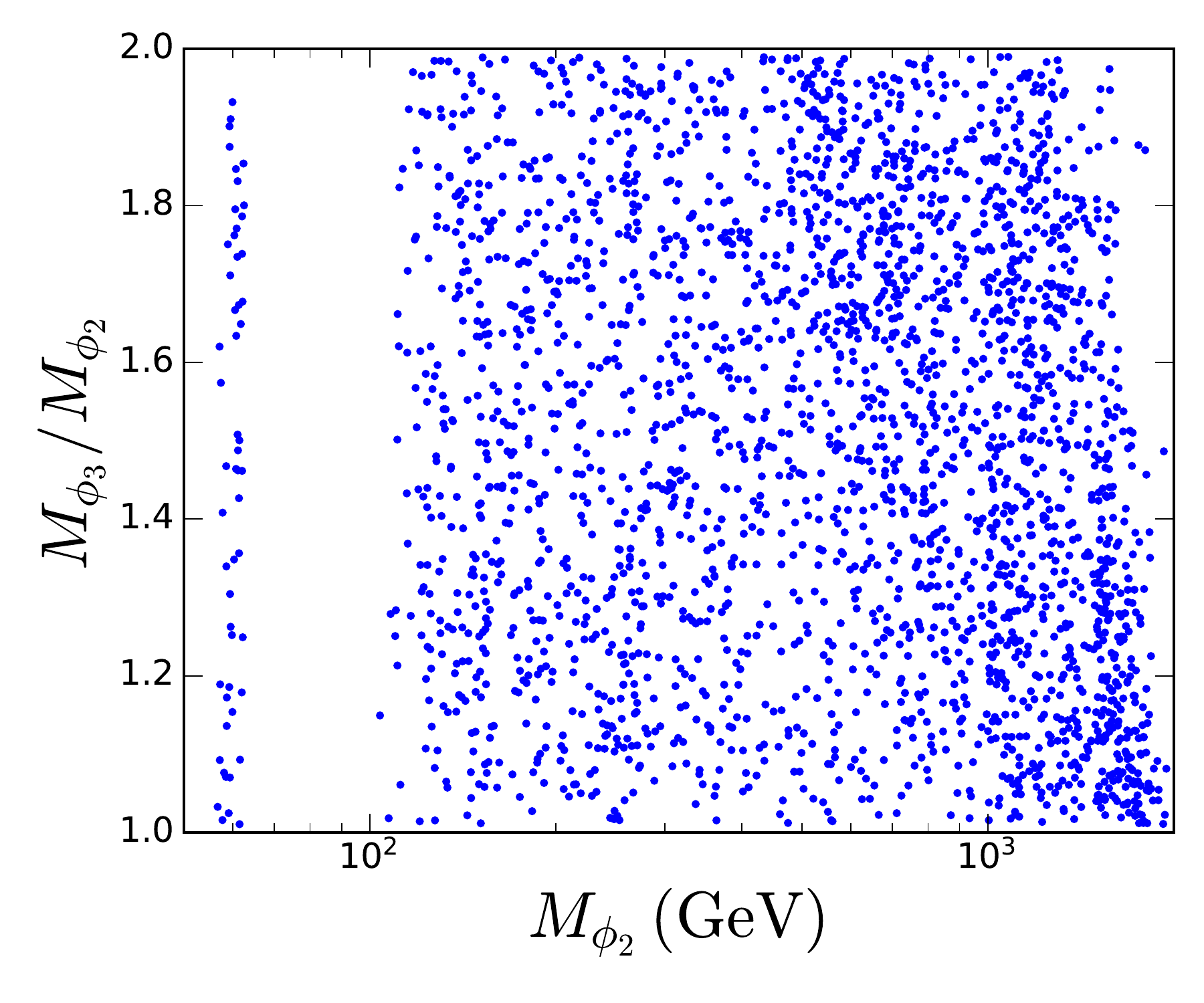}
\includegraphics[scale=0.44]{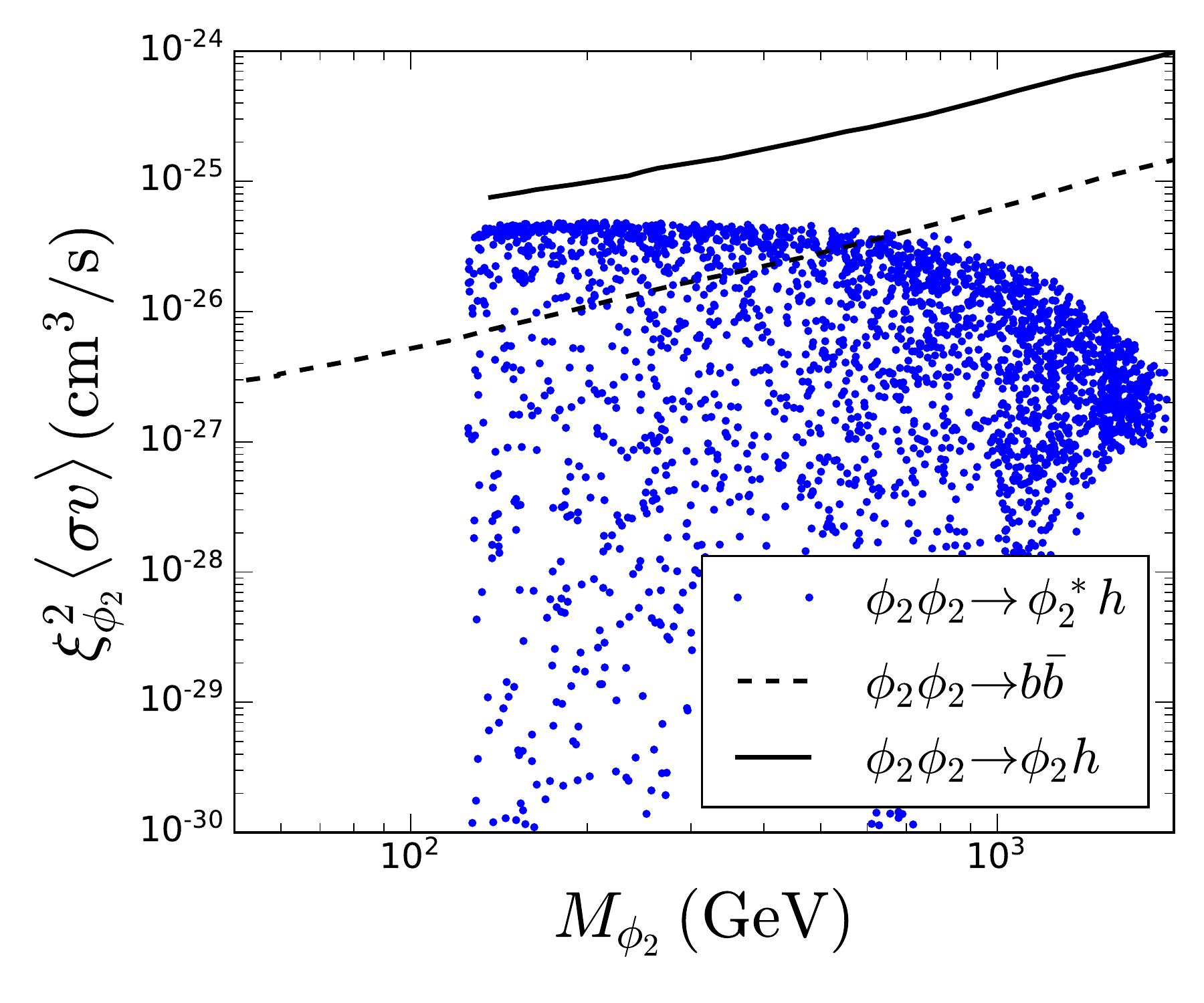}\\
\includegraphics[scale=0.44]{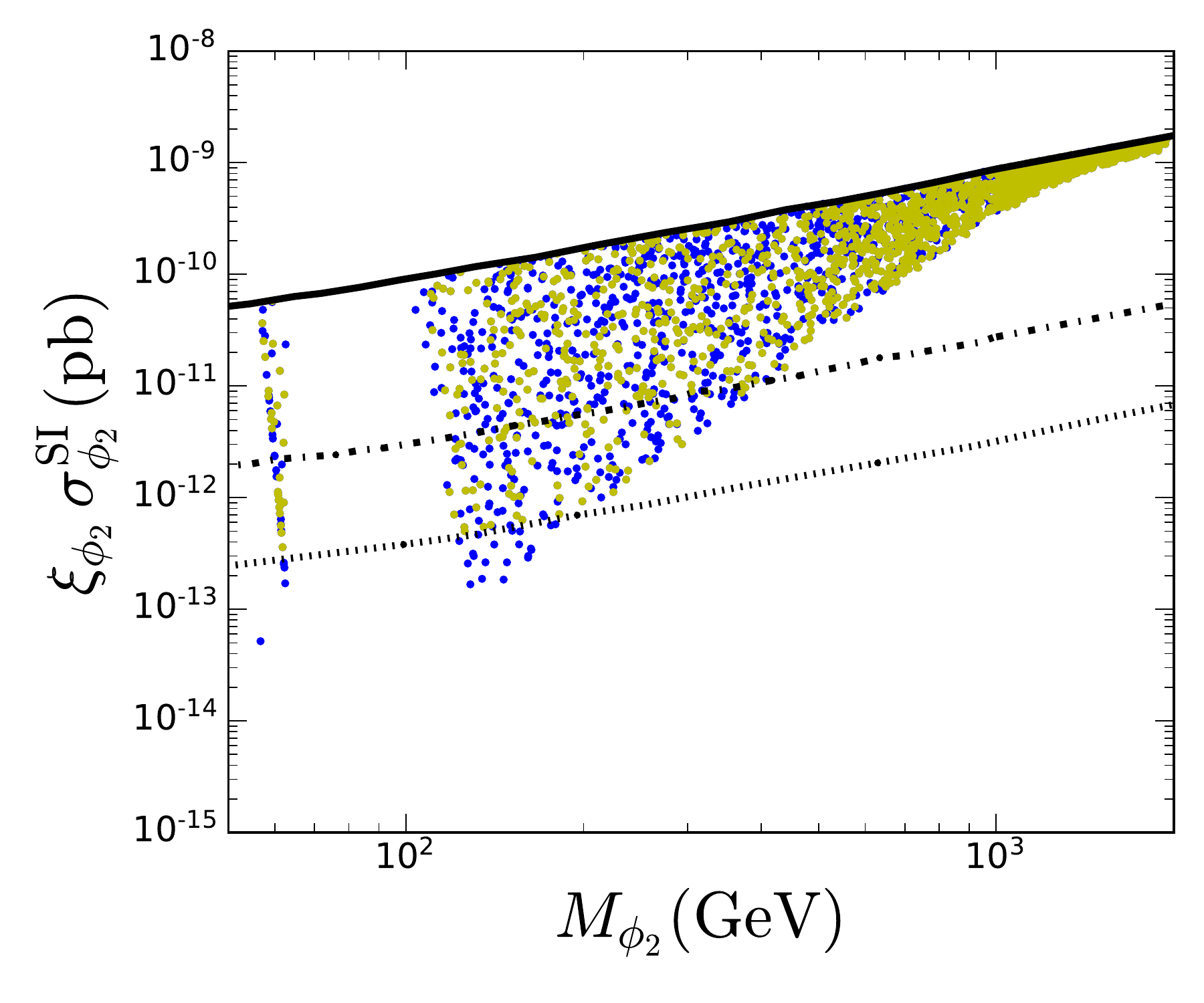}
\includegraphics[scale=0.44]{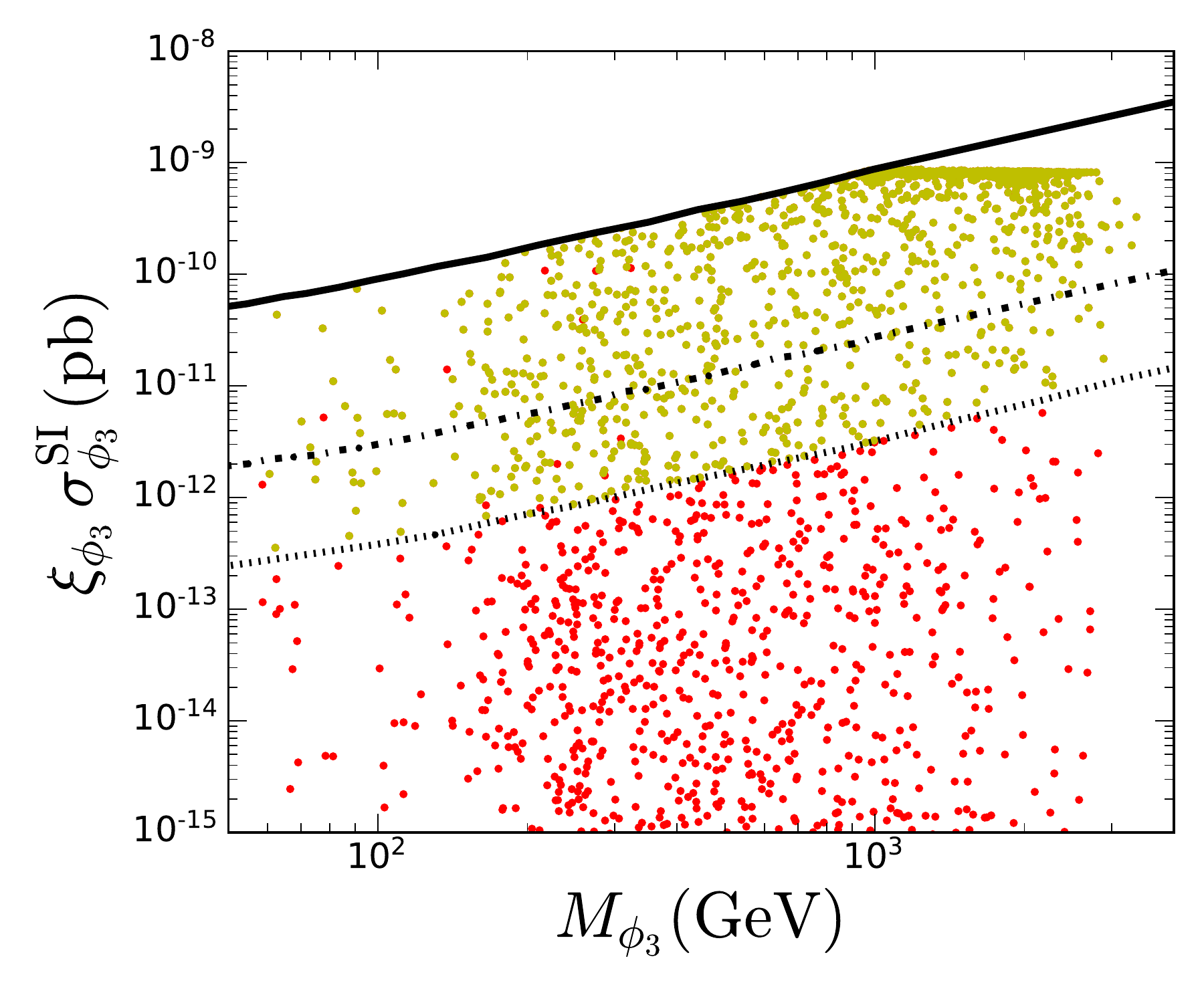}
\caption{A sample of viable points of the $Z_6(23)$ model with $M_{\phi_2}<M_{\phi_3}$, projected along different dimensions. The top panels show scatter plots of $M_{\phi_2}$ versus $\Omega_{\phi_2}/\Omega_{DM}$ (left) and versus $\mu_{32}$ (right). In the center panels, the ratio of  dark matter masses (left) and the most relevant indirect detection signal (right) are illustrated. The direct detection prospects are shown in the bottom panels for $\phi_2$ (left) and $\phi_3$ (right). }
\label{fig:scanZ623p2}
\end{figure}

For $M_{\phi_3}<M_{\phi_1}$, instead, there are no new viable points in our sample --the whole range of $M_{\phi_3}$ from 70 to 950 GeV  appears excluded. The reason is that the  process $\phi_1+\phi_3\to \phi_1+\phi_1$, which reduces the $\phi_3$ number density, is Boltzmann suppressed in this case.  This example illustrates the fact that the new interactions allowed by the $Z_{2n}$ symmetry not always succeed in opening up new regions of parameter space. In the next section, we will propose a simple extension of this scenario that succeeds in doing so. 

\subsection{$Z_6(23)$ model}
\begin{figure}
\centering
\includegraphics[scale=0.44]{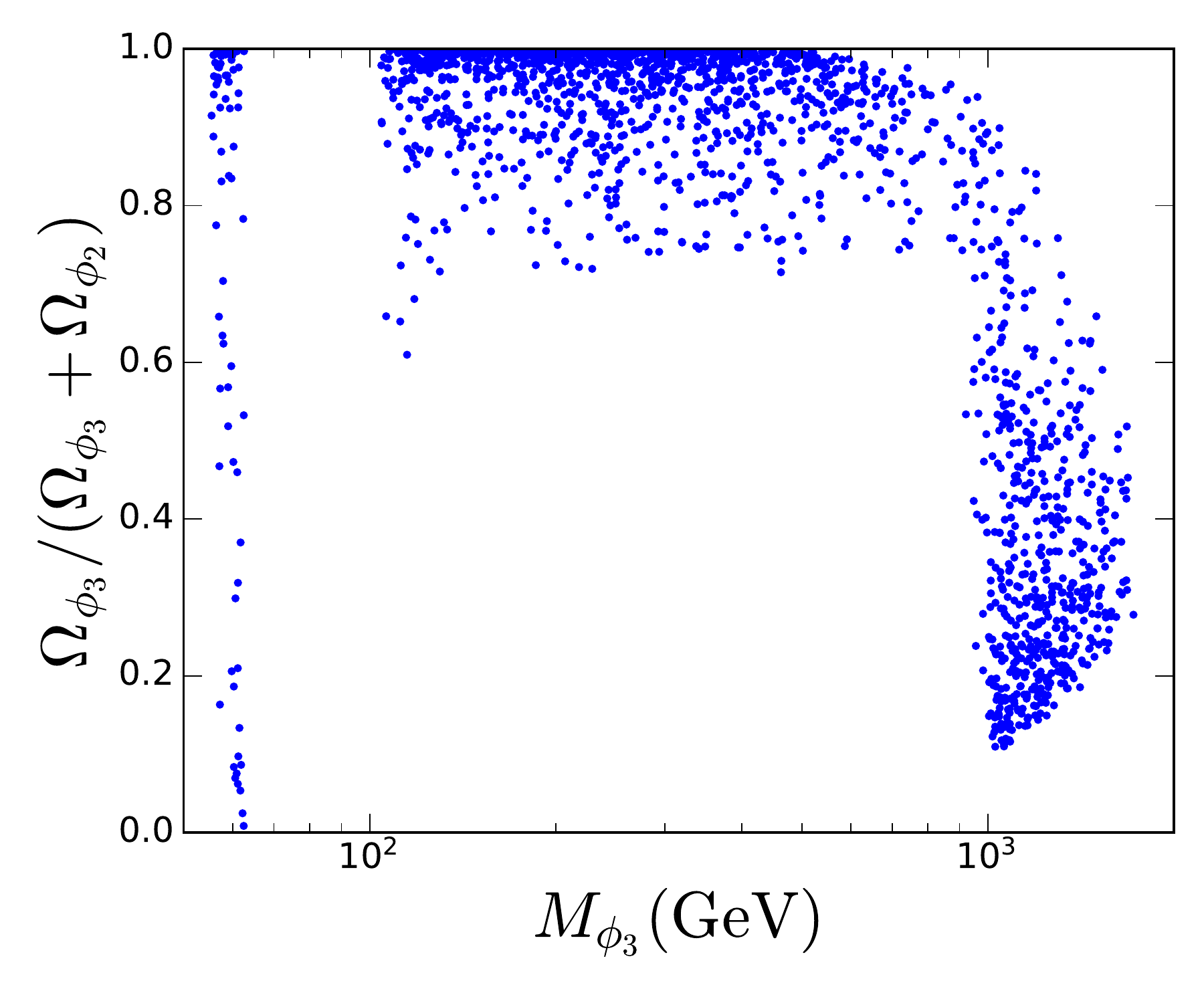}
\includegraphics[scale=0.44]{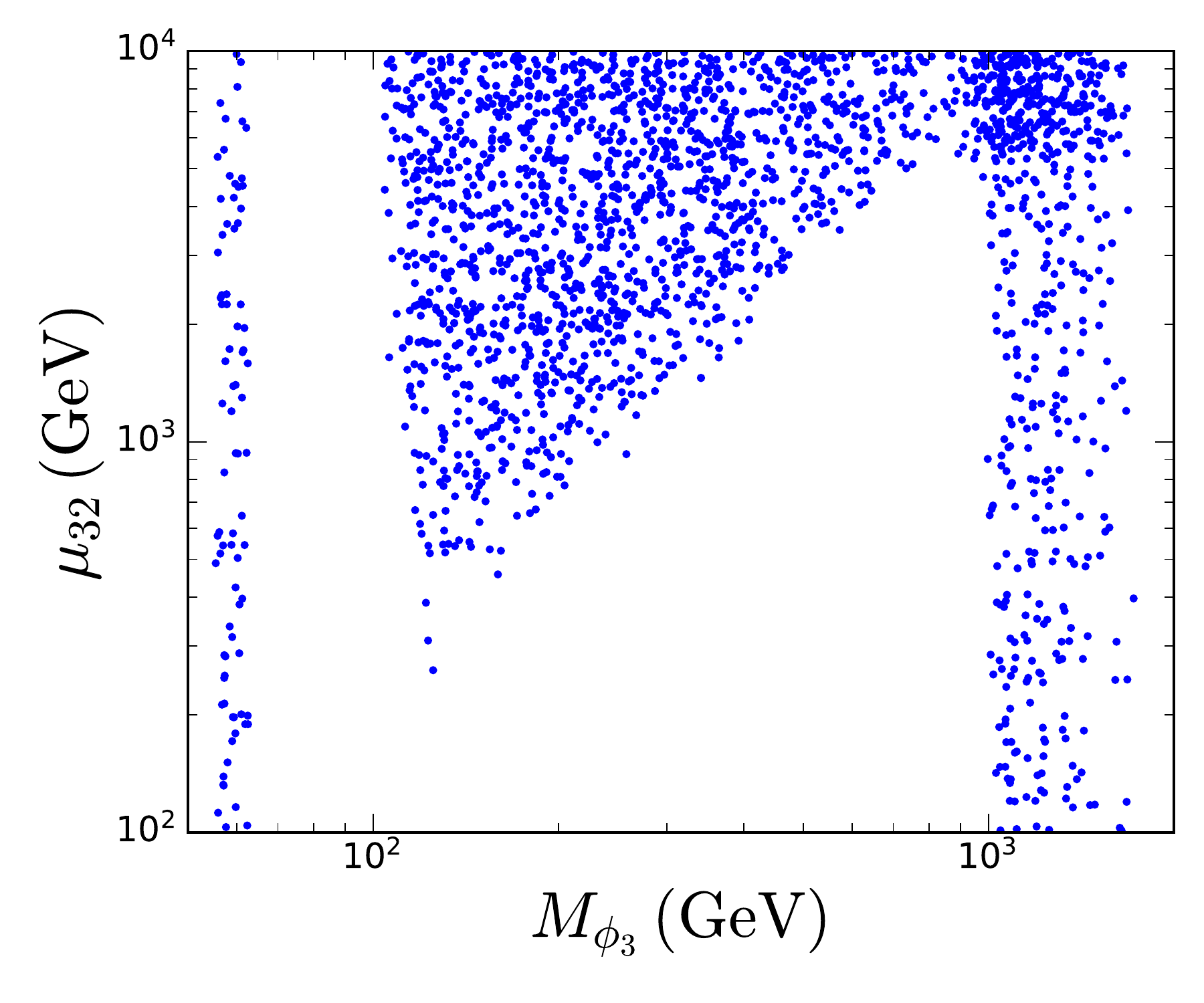}\\
\includegraphics[scale=0.44]{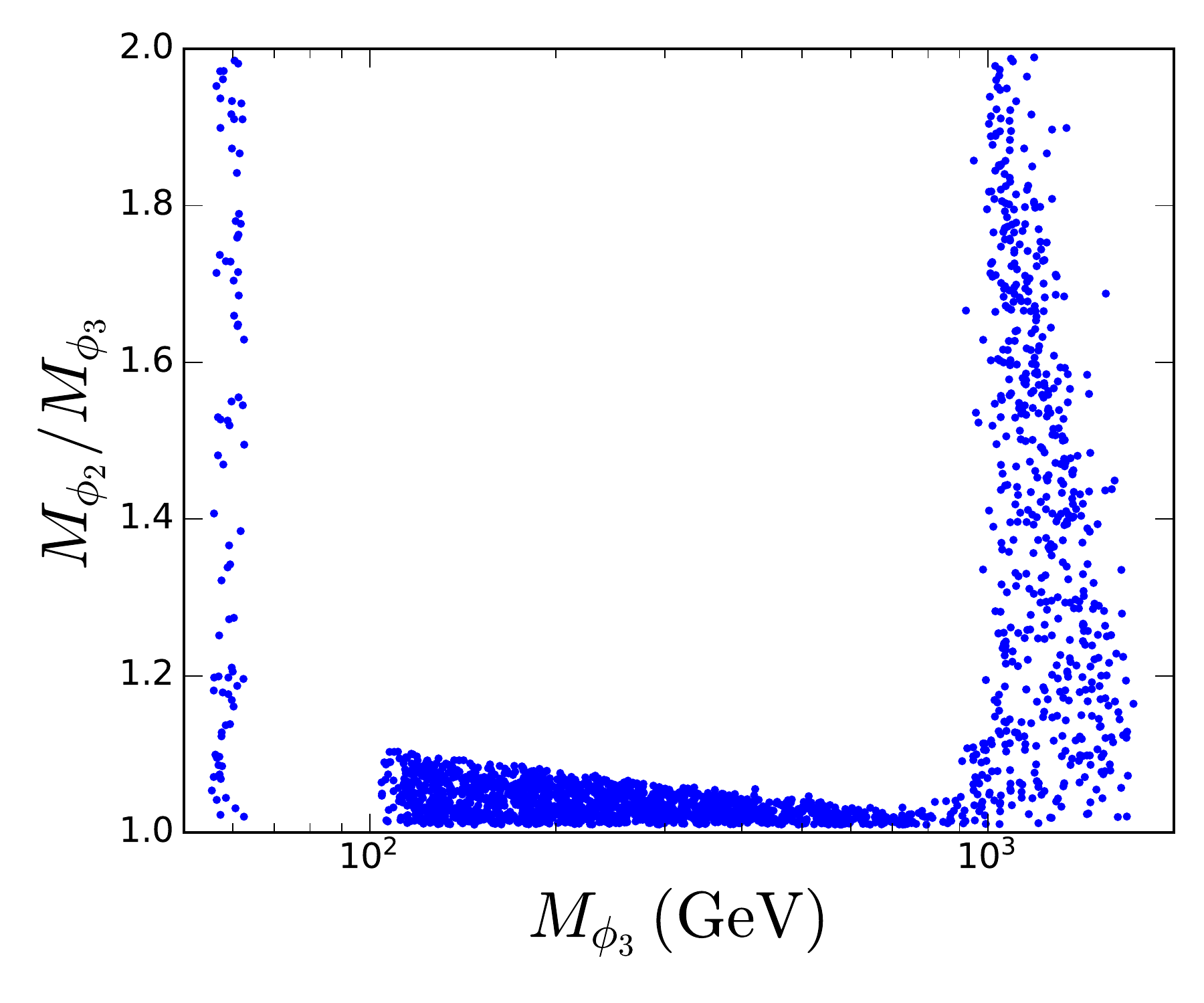}
\includegraphics[scale=0.44]{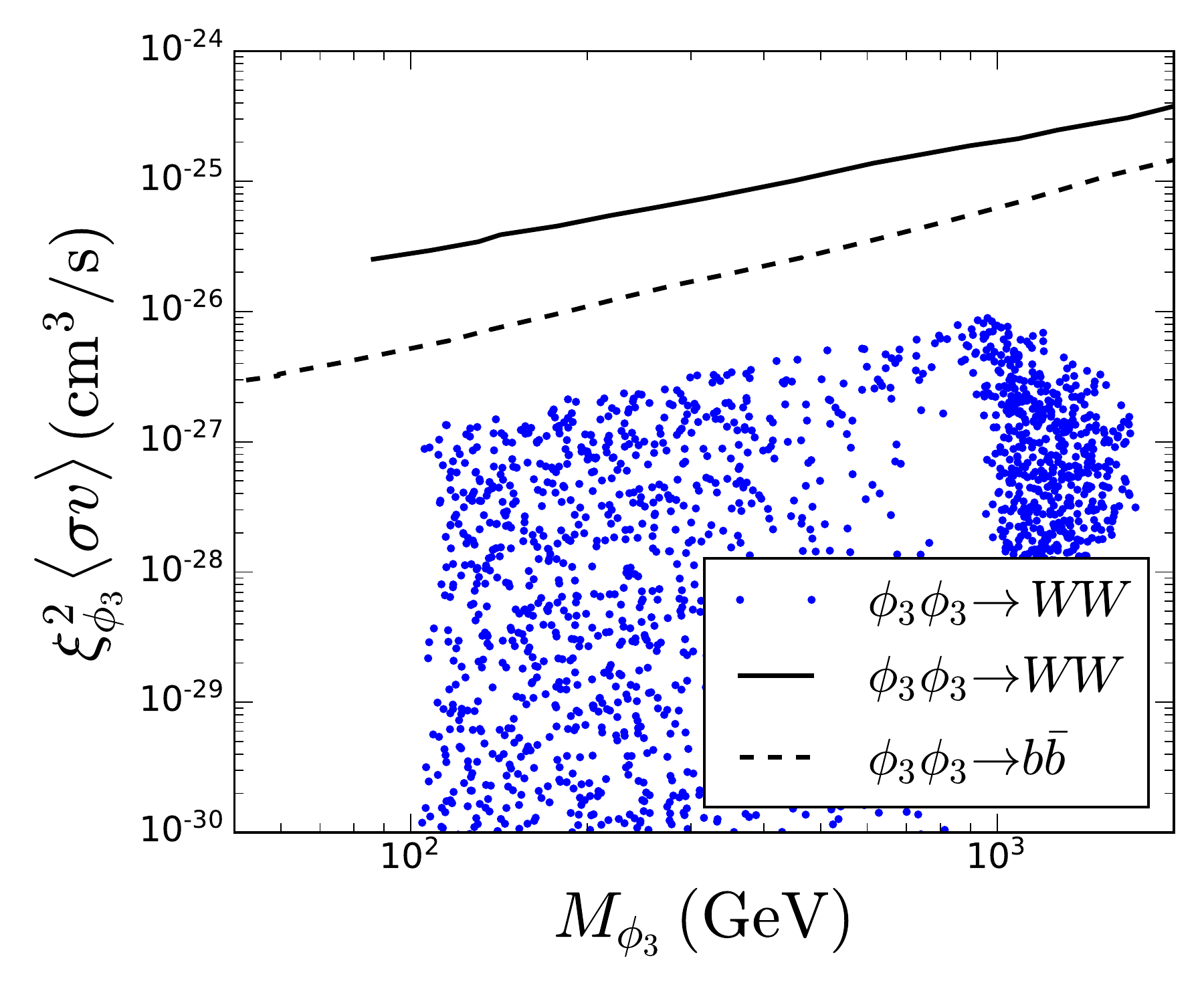}\\
\includegraphics[scale=0.44]{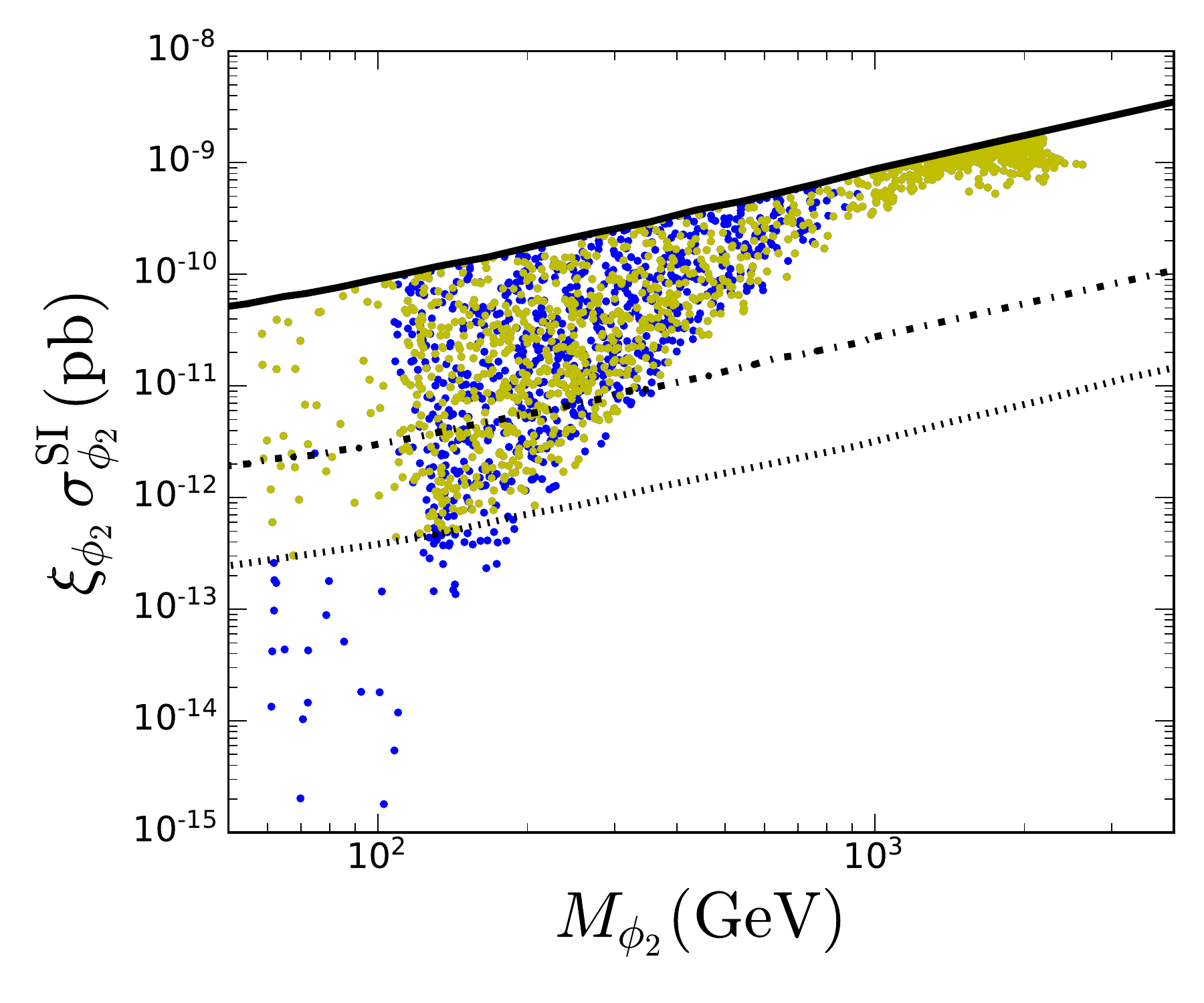}
\includegraphics[scale=0.44]{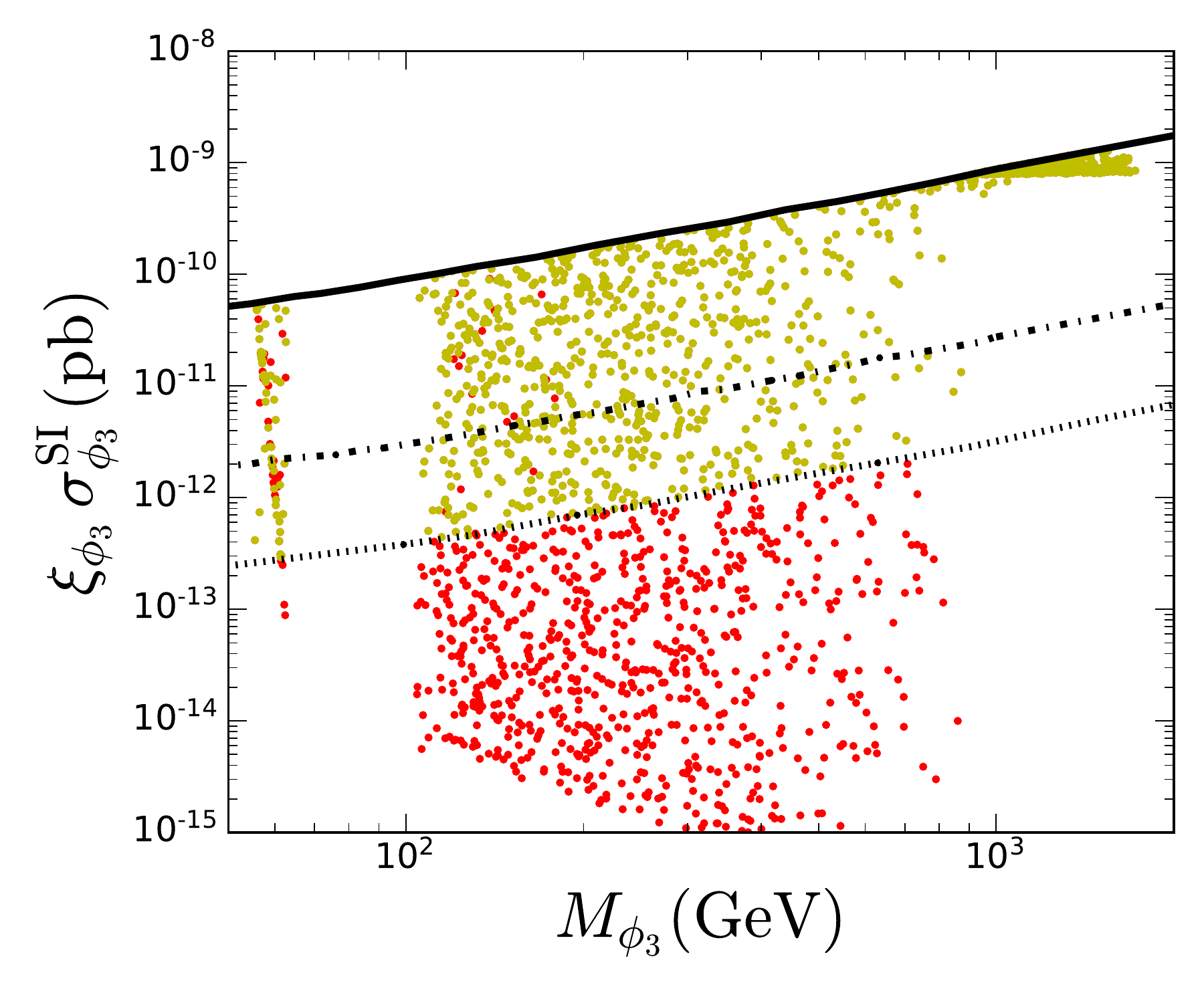}\\
\caption{A sample of viable points of the $Z_6(23)$ model with $M_{\phi_3}<M_{\phi_2}$, projected along different dimensions. The top panels show scatter plots of $M_{\phi_3}$ versus $\Omega_{\phi_3}/\Omega_{DM}$ (left) and versus $\mu_{32}$ (right). In the center panels, the ratio of  dark matter masses (left) and the most relevant indirect detection signal (right) are illustrated. The direct detection prospects are shown in the bottom panels for $\phi_2$ (left) and $\phi_3$ (right).}
\label{fig:scanZ623s1}
\end{figure}

The free parameters in this scenario are varied randomly using a logarithmic scan within the ranges
\begin{align}
    10^{-4}\leq & |\lambda_{423}|\leq 1\\
    100\,{\rm GeV}\leq & \mu_{32}\leq 10\,{\rm TeV}. 
\end{align}

For $M_{\phi_2}<M_{\phi_3}$, the results of the scan are displayed in figure \ref{fig:scanZ623p2}. Thanks to the semi-annihilation processes (see figure \ref{fig:semi3}),  the entire range above the higgs mass turns out to be viable for $M_{\phi_2}$ (and $M_{\phi_3}$). For $M_{\phi_2}\lesssim 1~\mathrm{TeV}$, no preference is observed in our sample regarding the fraction of the dark matter contributed by each particle, but for $M_{\phi_2}\gtrsim 1~\mathrm{TeV}$, both particles tend to contribute significantly  --see the top-left panel. In the top-right panel one can see that the minimum value of $\mu_{32}$ increases with $M_{\phi_2}$, as expected for a semi-annihilation driven relic density. With respect to the ratio of the dark matter masses, the center panel shows that it varies over a wide range for the viable points we found, indicating that the dark matter particles are not required to be degenerate, in stark contrast with the $Z_6(13)$ model. The dominant dark matter annihilation channel in our sample happens to be the semi-annihilation process $\phi_2+\phi_2\to \phi_2^*+h$. And, as seen from the center-right panel,  indirect detection does not currently constraint our set of viable models. A remarkable feature of this scenario is that the detection of $\phi_2$ at DARWIN is practically guaranteed for all the points in our sample --only few points at low masses may evade detection in such an experiment (see bottom-left panel). Such an encouraging situation had not arisen in the previous scenarios we discussed.

For $M_{\phi_3}<M_{\phi_1}$, see figure \ref{fig:scanZ623s1},   the  range of $M_{\phi_3}$ with viable points  goes up to 1700 GeV,  where $\lambda_{S_3}$ saturates the maximum value considered in our scan.  The viable points below $M_{\phi_3}\lesssim 950$ GeV demand a strong mass degeneracy, namely $M_{\phi_2}/M_{\phi_3}\lesssim 1.1$ (see center panel), and a sizeable $\lambda_{423}$ quartic coupling, $|\lambda_{423}|\gtrsim0.1$.  Indeed,  the semi-annihilation processes generated by the $\mu_{32}$ interaction only affect the number density of the $\phi_2$, which means that $\phi_3$ annihilates through the Higgs portal interactions or Boltzmann-suppressed conversion processes. The dominant annihilation channel in our sample is now $\phi_3+\phi_3\to WW$ but its rates are too small to be observed with current or future Fermi data --see center-right panel. The direct detection prospects, on the contrary, are seen to be similar to the previous case (see bottom panels), with an almost certain detection of $\phi_2$ at DARWIN, but with the caveat that the dark matter particles are highly degenerate, implying that it will be  challenging to distinguish this scenario from those with just one dark matter particle.

In this section we investigated the phenomenology of three scenarios for two-component dark matter based on the $Z_4$ and $Z_6$ symmetries. In all of them the dark matter consists of two scalars that are singlets under the SM gauge group --one of them is complex and the other is real.  Our analysis reveals that, thanks to the new semi-annihilation and dark matter conversion processes that are allowed by the $Z_{2n}$ symmetries, it is often possible to satisfy all the theoretical and experimental bounds over a wide range of dark matter masses, especially below 1 TeV.   These models are also testable via current and forthcoming direct detection experiments, which may be able to detect signals from both dark matter particles.

\section{Non-minimal models}\label{sec:extension}
As we have seen in the previous section, in some cases the new $Z_{2n}$ interactions are unable to open up new regions of viable parameter space. It is natural to ask, therefore, whether there is a simple way to extend such models and achieve that goal.  A direct and rather trivial extension of the framework assumed so far is to consider $\phi_B$ as a complex field. In this section we analyze such a possibility. Let us stress that unlike $\phi_A$, which is necessarily complex, $\phi_B$ may be real, but it does not have to be so. We will see that when $\phi_B$ is complex, it splits into two real fields, the lighter one being a dark matter particle (the other dark matter particle is $\phi_A$).  We call these models non-minimal because they contain an extra particle in the dark sector (the heavier one left from the splitting of $\phi_B$) that is unstable and does not contribute to the dark matter.  

\begin{figure}[t]
\centering
\includegraphics[scale=0.8]{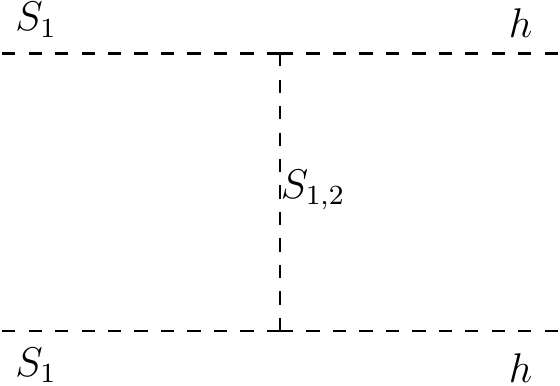}\hspace{0.4cm}
\includegraphics[scale=0.8]{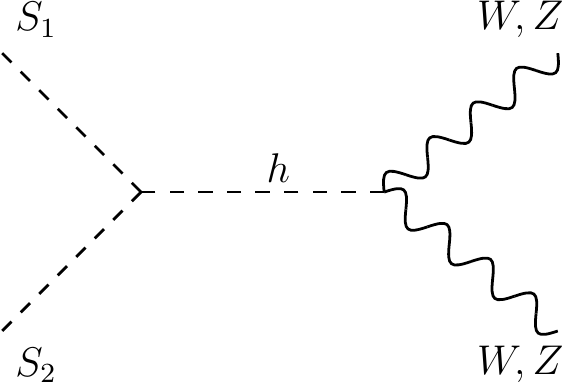}\hspace{0.4cm}
\includegraphics[scale=0.8]{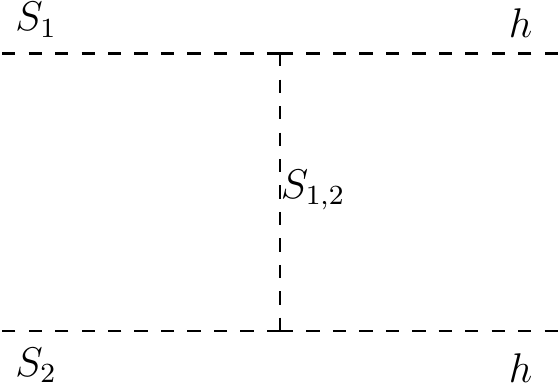}\hspace{0.4cm}
\caption{Dark matter $S_1$ annihilation (left) and $S_1-S_2$ coannihilation (center and right) processes involving the $\lambda_{6}$ and $\lambda_{SB}$ interactions.}
\label{fig:coanni-phiB-complex}
\end{figure}

Under the assumption that $\phi_B$ is complex the corresponding potential can be written as the sum of three contributions,
\begin{align}
 \mathcal{V}_{Z_{2n}}(\phi_A,\phi_B)&=\,\mathcal{V}_{1}(\phi_A,\phi_B)+\mathcal{V}_2(\phi_A,\phi_B)+\mathcal{V}_3(\phi_A,\phi_B), 
 \end{align} 
where the first and second contributions are analogous to the ones in the case of $\phi_B$ real. Concretely, for $\mathcal{V}_1$ we have
\begin{align}\label{eq:V1complex}
 \mathcal{V}_1(\phi_A,\phi_B)&\equiv\,\,\mu^2_H|H|^2+\lambda_H|H|^4+\mu_{A}^2|\phi_A|^2+\lambda_{4A}|\phi_A|^4+\mu_{B}^2|\phi_B|^2+\lambda_{4B}|\phi_B|^4\nonumber\\
  & \,+\lambda_{4AB}|\phi_A|^2|\phi_B|^2+\lambda_{SA}|H|^2|\phi_B|^2+\lambda_{SB}|H|^2|\phi_B|^2.
 \end{align}
In the third contribution, the additional terms associated to the field $\phi_B$ that are invariant for any $Z_{2n}$ symmetry are grouped. It can be written as
 \begin{align}
\mathcal{V}_3(\phi_A,\phi_B)&\equiv\, \frac{1}{2}\left[\kappa^2 \phi_B^2 + \lambda_{6}|H|^2 \phi_B^2 + \lambda_{5B} \phi_B^4 +\lambda_{4AB}' |\phi_A|^2\phi_B^2 + \lambda_{5B}' |\phi_B|^2\phi_B^2 \right]+\text{h.c.}. 
 \end{align}
After the electroweak symmetry breaking the real and imaginary components of $\phi_B=(\phi_{BR}+i\phi_{BI})/\sqrt{2}$ get mixed through the $\kappa^2$ and $\lambda_6$ terms in $\mathcal{V}_3$. Specifically, the $2\times2$ mass matrix in the basis  $(\phi_{BR},\phi_{BI})$ has the entries 
\begin{align}
    M^2_{11}&=\mu_B^2+\Re(\kappa^2)+\frac{1}{2}(\lambda_{SB}+\Re(\lambda_6))v_H^2,\\
    M^2_{22}&=\mu_B^2-\Re(\kappa^2)+\frac{1}{2}(\lambda_{SB}-\Re(\lambda_6))v_H^2,\\
    M^2_{12}&=M^2_{21}=-\Im(\kappa^2)- \frac{1}{2}\Im(\lambda_6) v_H^2.
\end{align}
Notice that $M_{12}^2$ is in general nonzero because it is not possible to render $\kappa^2$ and $\lambda_6$ real by a field phase redefinition of $\phi_B$.  The mass eigenstates $S_1,S_2$ are defined through the rotation matrix
\begin{align}
\begin{pmatrix}
    \phi_{BR} \\
    \phi_{BI}
\end{pmatrix}&=
\begin{pmatrix}
    \cos\theta & \sin\theta \\
    -\sin\theta & \cos\theta
\end{pmatrix}
\begin{pmatrix}
    S_1 \\
    S_2
\end{pmatrix},
\end{align}
with a mixing angle 
\begin{align}
    \sin2\theta&=\frac{2\Im(\kappa^2)+\Im(\lambda_6)v_H^2}{M_{S_1}^2-M_{S_2}^2},
\end{align}
and masses
\begin{align}
    M_{S_1,S_2}^2&=\frac{1}{2}\left[2\mu_B^2+\lambda_{SB}v_H^2\mp\sqrt{\left[2\Im(\kappa^2)+\Im(\lambda_6)v_H^2\right]^2+\left[2\Re(\kappa^2)+\Re(\lambda_6)v_H^2\right]^2}\right],
\end{align}
where we assumed, without loss of generality, that $S_1$ is the lightest of the mass eigenstates, $M_{S_1}<M_{S_2}$. Let us stress that the existence of complex parameters in $\mathcal{V}_3$ is required to have a mixing between $\phi_{BR}$ and $\phi_{BI}$.

From $\mathcal{V}_3$, the trilinear interaction between $S_1$, $S_2$ and the Higgs boson can be seen to be
\begin{align} \label{eq:s1s2h}
  \mathcal{L}&\supset v_H\lambda_6\sin 2\theta S_1S_2h,
\end{align}
which implies that, as long as $\lambda_6$ and $\theta$ are non-zero, $S_2$ is unstable independently of its mass.
In addition, the interaction in equation (\ref{eq:s1s2h}) leads to coannihilation processes between $S_1$ and $S_2$ mediated by the Higgs boson (see figure \ref{fig:coanni-phiB-complex}). Such processes are absent in $Z_{2n+1}$ models and constitute  a novelty of the scenarios in this section. 

\begin{figure}[t]
\centering
\includegraphics[scale=0.44]{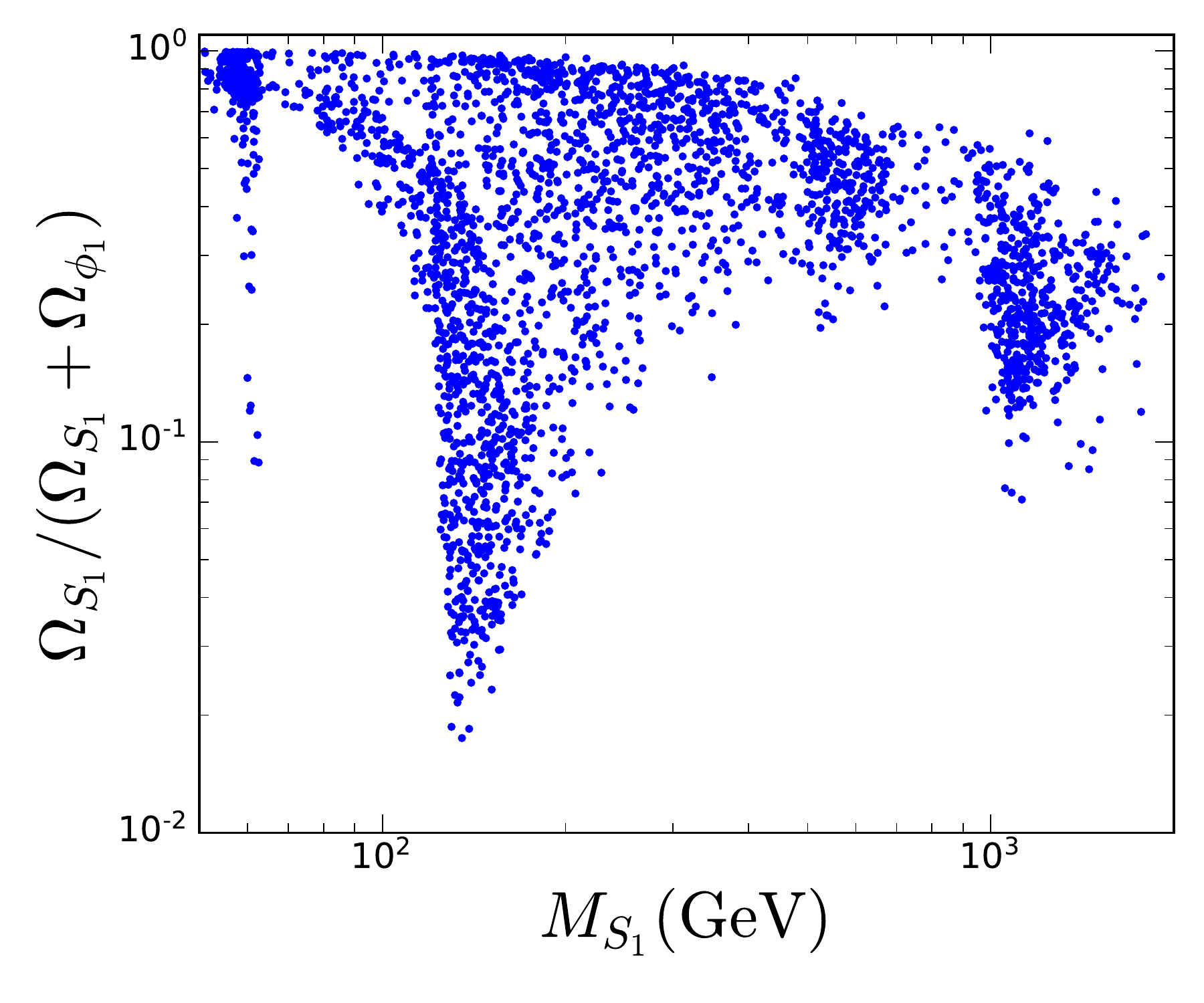}
\includegraphics[scale=0.44]{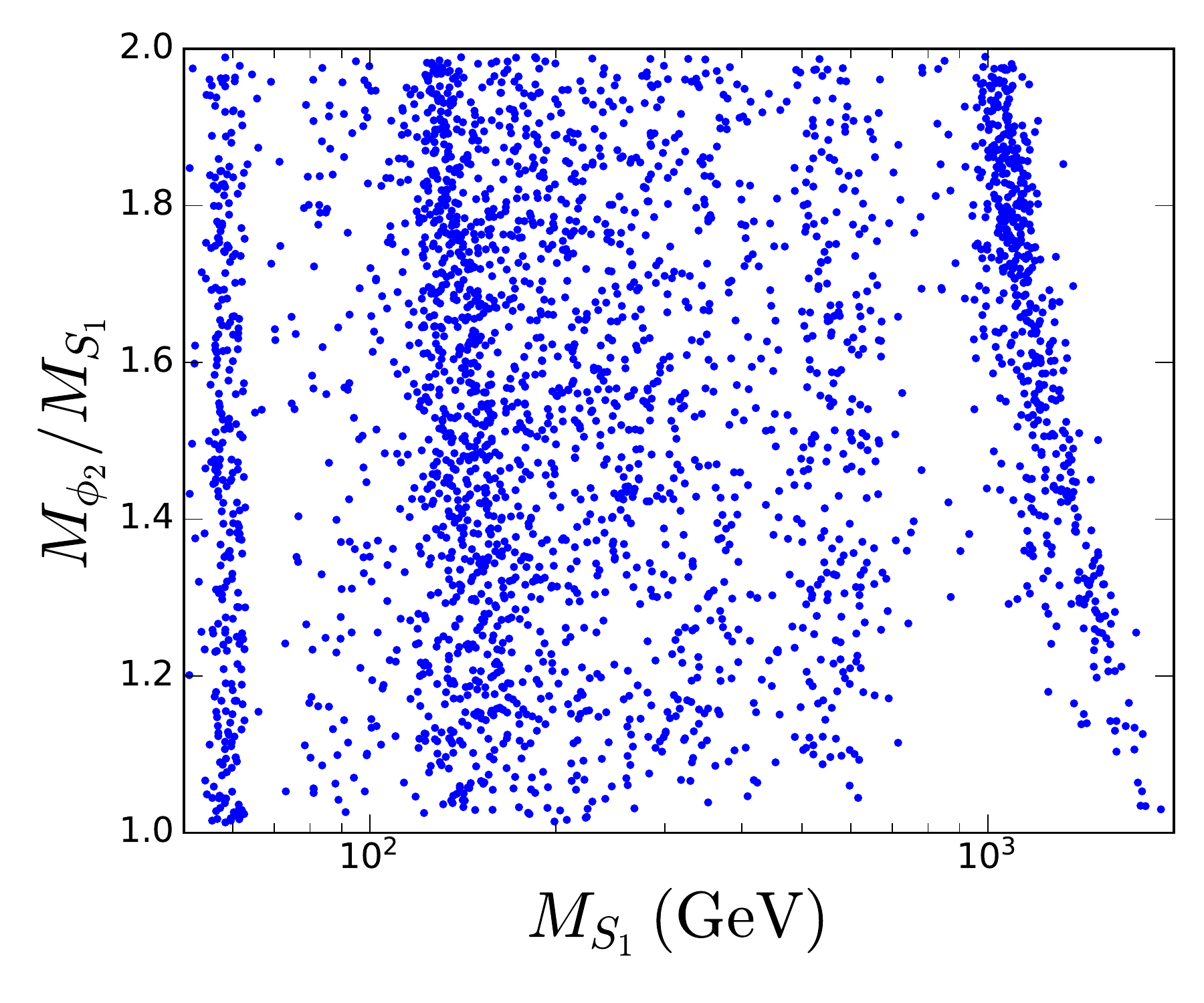}\\
\includegraphics[scale=0.44]{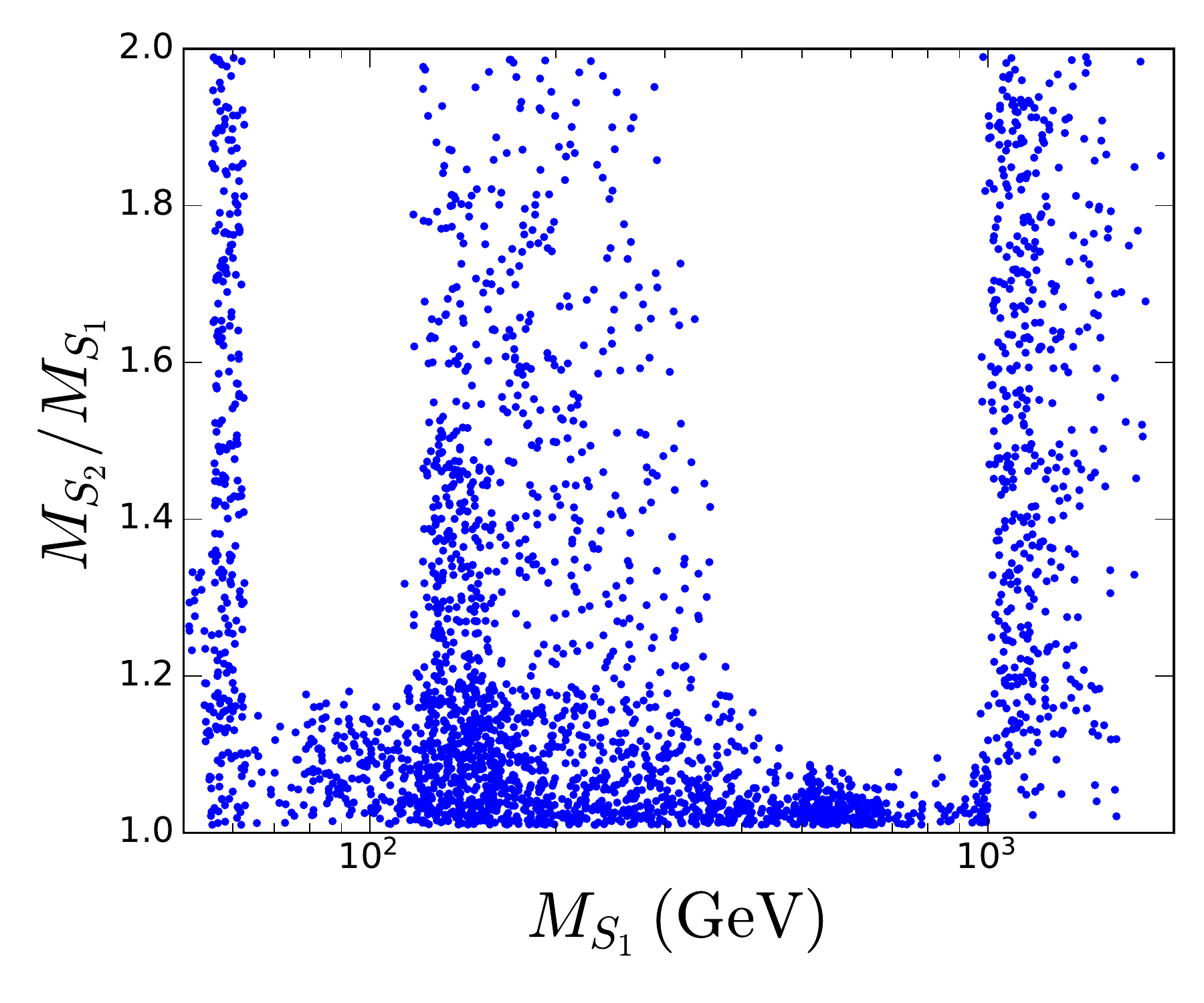}\\
\includegraphics[scale=0.44]{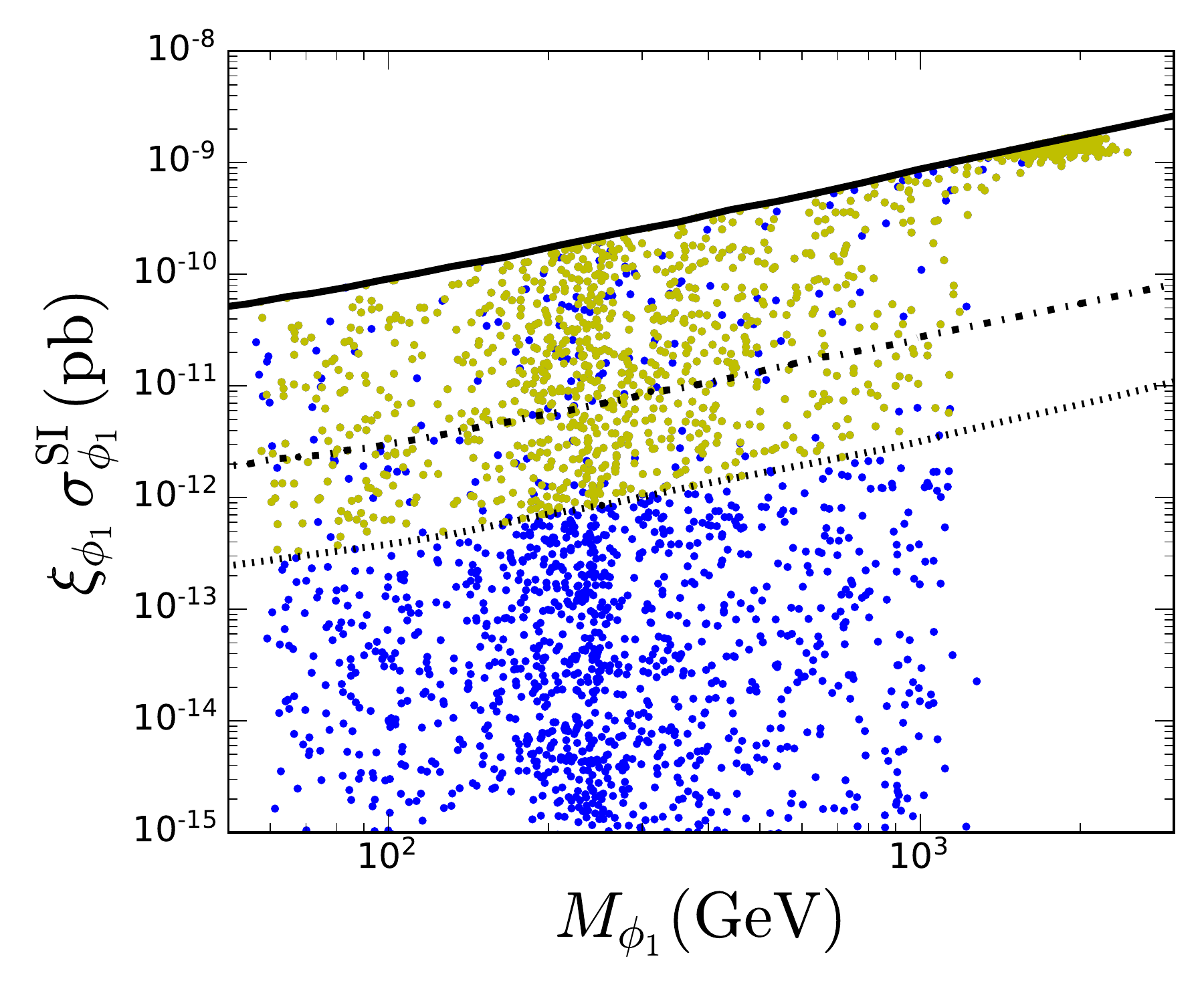}
\includegraphics[scale=0.44]{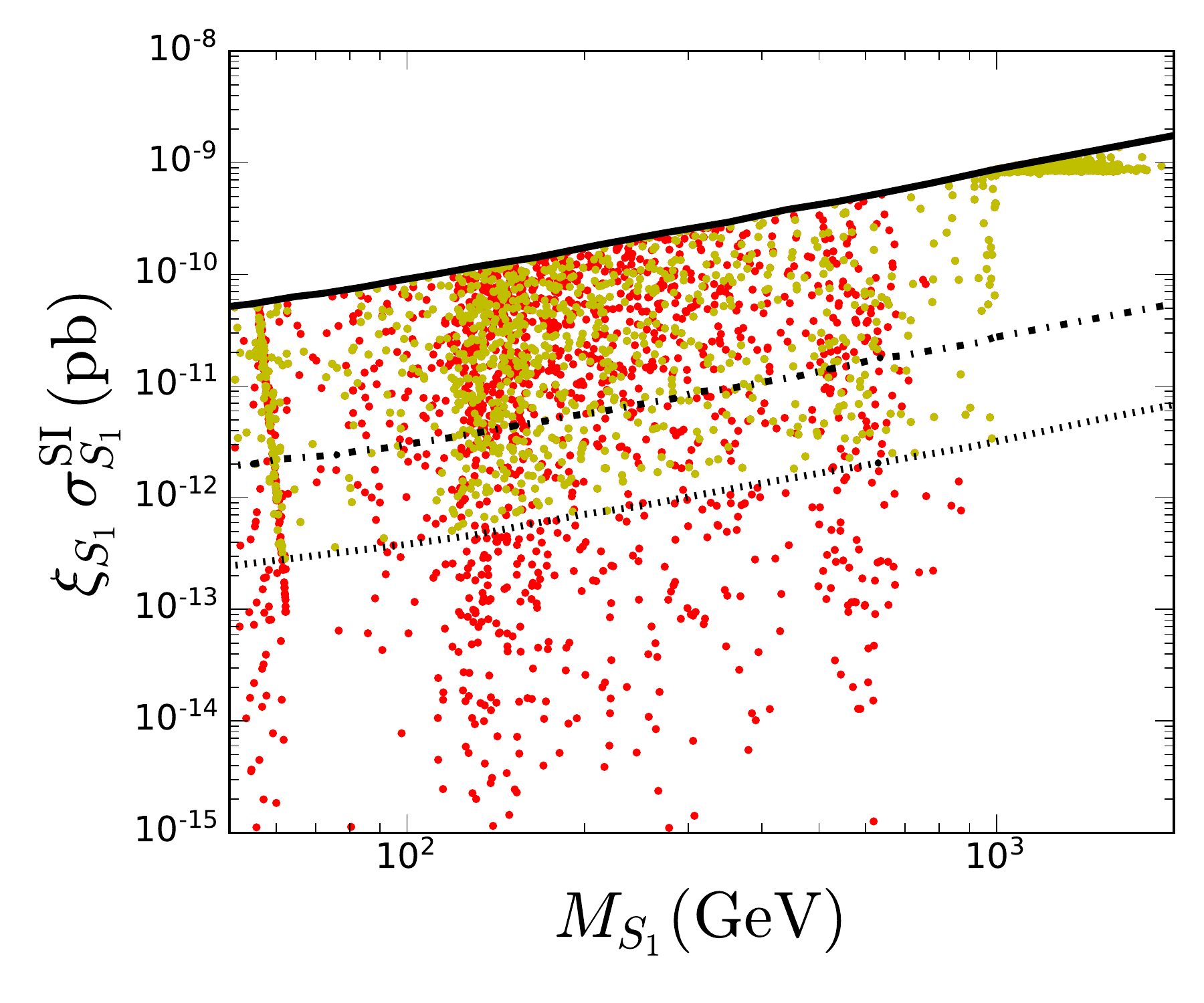}
\caption{A sample of viable points of the $Z_6(13)$ model with complex fields and $M_{S_1}<M_{\phi_1}$, projected along different dimensions. The top panels show scatter plots of $M_{S_1}$ versus $\Omega_{S_1}/\Omega_{DM}$ (left) and versus the ratio of  dark matter masses  (right). The center panel show the ratio $M_{S_2}/M_{S_1}$ --see text for details.  The direct detection prospects are shown in the bottom panels for $\phi_1$ (left) and $S_1$ (right).}
\label{fig:scanZ613-S1}
\end{figure}

As a case study, let us  extend the $Z_6(13)$ scenario along these lines.  Accordingly, the scalar potential $\mathcal{V}_2$ now reads 
\begin{align}\label{eq:V2Z613-complex}
 \mathcal{V}_{2}^{Z_6}(\phi_1,\phi_3)&=\frac{1}{3} \lambda_{41}'\phi _1^3\phi_3 + \frac{1}{3} \lambda_{42}'\phi_1^3\phi_3^* + \text{h.c.}.
\end{align} 
To establish the impact of considering $\phi_3$ complex, the magnitude of the relevant quartic free parameters of this scenario are varied between $10^{-4}$ and $1$. 
The results from  the scan are displayed in figure \ref{fig:scanZ613S1} for the $M_{S_1}<M_{\phi_1}$ case (the results for the $M_{\phi_1}<M_{S_1}$ case are the same of those with $\phi_3$ real). 
This case admits  $S_1S_2$ coannihilations,  which are quite efficient on depleting $\Omega_{S_1}$ and therefore the whole range for $M_{S_1}$ remains valid, with the remarkable fact that $\Omega_{S_1}$ can be the subdominant contribution for $M_{S_1}\gtrsim100$ GeV, reaching even values less than $10\%$ of the total dark matter relic density.  
The results also indicate that both $\phi_1$ and $S_1$ may be observed at current and forth-coming direct detection experiments, as it is shown in the bottom panels. 

It is clear then that by taking $\phi_3$ complex rather than real  in the $Z_6(13)$ model, new viable regions of parameter space have appeared, and the model has become viable for dark matter masses below $1$ TeV. This kind of extension can straightforwardly be applied to other $Z_{2n}$ models.

\section{Discussion}\label{sec:discussion}
The results presented in the previous sections make manifest the fact that the specific interactions associated to each $Z_{2n}$ symmetry ($\mu_{S_1}$, $\lambda'_{41}$ and $\mu_{32}$ for the $Z_4$, $Z_6(13)$ and $Z_6(23)$ scenarios, respectively) are crucial in reducing the number density of the lighter dark matter particle.  Without them, there would be no new viable regions of parameter space.  We can now extrapolate what we have learned about these scenarios to qualitatively discuss  the dark matter phenomenology of other $Z_{2n}$ models.  

Let us  denote the two dark matter particles by  $\phi_A$ (the complex one) with $A<n$ and $\phi_B=\phi_n$ (the real one) with $Z_{2n}$ charges $w^A$ and $w^n=-1$, respectively. For a $Z_8$ ($n=4$) symmetry the  complete set of possibilities for the two dark matter particles are $A=1,2,3$.
For the scenarios $(\phi_1,\phi_4)$ and $(\phi_3,\phi_4)$ the $Z_8$ symmetry does not allow any extra invariant term in $\mathcal{V}_2$, that is  
\begin{align}
    \mathcal{V}_2^{Z_8}(\phi_1,\phi_4)=\mathcal{V}_2^{Z_8}(\phi_3,\phi_4)=0.  
\end{align}
Thus we expect, in analogy with the  $Z_6(13)$ scenario already discussed, no viable points for  $70 \lesssim M_{\phi_1}/\mathrm{GeV}\lesssim 1850$  in the case $M_{\phi_1}<M_{\phi_3}$ and in the range $70 \lesssim M_{\phi_3}/\mathrm{GeV}\lesssim 950$ in the case $M_{\phi_3}<M_{\phi_1}$. The third possibility is the scenario $(\phi_2,\phi_4)$ where the $Z_8$ symmetry allows terms that are equivalent to Eq. (\ref{eq:V3Z4}), that is, 
  \begin{align}
    \mathcal{V}_2^{Z_8}(\phi_2,\phi_4)&=\, \frac{1}{2}\left[\mu_{S2}\phi^2_2\phi_4 + \lambda_{52} \phi_2^4\right]  + \text{h.c.}.
  \end{align}
Therefore the results obtained for the $Z_4$  model are the same as for this scenario. And this conclusion can be generalised to all the scenarios based on a $Z_{4n}$ symmetry with dark matter fields $\phi_{n}$ and $\phi_{2n}$.

Along the same lines,  all the scenarios with $n=5$ feature 
\begin{align}
\mathcal{V}_2^{Z_{10}}(\phi_A,\phi_5)=0,     \hspace{1cm}A=1,2,3,4,
\end{align}
so that again no new viable regions are expected. Scenarios invariant under a $Z_{6n}$ symmetry with dark matter fields $\phi_{n}$ and $\phi_{3n}$  ($\phi_{2n}$ and $\phi_{3n}$) turn out to be  equivalent to the $Z_6(13)$ ($Z_6(23)$) scenario we already analyzed. All these results suggest that the $Z_4$ and $Z_6$ models  studied in this work constitute prototypes for all the $Z_{2n}$ models with one complex and one real fields.

Finally, let us recall that  we have identified the parameters  that are expected to be relevant for the different models, and have illustrated the generic mechanisms that allow to satisfy the different bounds imposed. In this sense, therefore, this work constitutes a first step towards the elaboration of deep statistical analysis involving different sampling algorithms~\cite{Martinez:2017lzg} such as  Markov chain Monte Carlo~\cite{Dunkley:2004sv} or multimodal nested sampling \cite{Feroz:2008xx}.

\section{Conclusions}\label{sec:conclusions}
In this paper we considered two-component dark matter scenarios with scalar singlets that transform non-trivially under a $Z_{2n}$ symmetry, where the dark matter consists of a complex field and a real field. We  payed particular attention to the $n=2$ and $n=3$ cases since they represent the simplest realizations of these scenarios and serve as prototypes for higher $Z_{2n}$ models.  
We performed a phenomenological analysis of three scenarios based on the $Z_4$ and $Z_6$ symmetries, and obtained using a logarithmically-uniform distribution) a large sample of points compatible with current bounds. The analysis of this sample shows that, thanks to the semi-annihilations induced by the trilinear interactions, the $Z_4$ and $Z_6(23)$ scenarios become viable over the entire range of dark matter masses considered, $M\lesssim 2~\mathrm{TeV}$, and that they can be probed by future direct detection experiments. In fact, both dark matter particles could be detected in several cases, and,  in the $Z_6(23)$ scenario,  the observation of one of them at DARWIN is practically  guaranteed. The $Z_6(13)$ scenario, on the other hand, is quite constrained because semi-annihilation processes are absent while  dark matter conversion processes are inefficient at depleting the abundance of the lightest dark matter particle. Nonetheless, by considering as complex both dark matter fields we showed that it is possible to revert substantially this result. All in all, we demonstrated that this new kind of two-component dark matter models are viable and interesting alternatives to explain one of the greatest mysteries of our time --the nature of the dark matter.

\section*{Acknowledgments}
The work of OZ is supported by Sostenibilidad-UdeA and the UdeA/CODI Grants 2017-16286 and 2020-33177.

\bibliography{references}

\end{document}